\documentclass{article}
\usepackage{latexsym}
\usepackage{amsmath}
\usepackage{amsfonts}
\usepackage{amssymb}

\newcommand{\ms}{\medskip}

\newcommand{\noi}{\noindent}
\newcommand{\ra}{\rightarrow}
\newcommand{\bea}{\begin{eqnarray}}
\newcommand{\eea}{\end{eqnarray}}
\newcommand{\ol}{\overline}
\newcommand{\gr}{Groenewold}
\newcommand{\vh}{Van~Hove}
\newcommand{\vn}{Von~Neumann}
\newcommand{\q}{{\cal Q}}
\newcommand{\p}{C^{\infty}(M)}

\newcommand{\ci}{C^{\infty}}
\newcommand{\h}{{\cal H}}
\newcommand{\f}{{\cal F}}

\newcommand{\A}{{\cal A}}

\newcommand{\oo}{{\cal O}}
\newcommand{\s}{{\cal S}}
\newcommand{\uu}{{\cal U}}
\newcommand{\fn}{{\mathfrak n}}
\newcommand{\fd}{finite-dimensional}
\newcommand{\id}{infinite-dimensional}
\newcommand{\fc}{finite-codimensional}
\newcommand{\pa}{Poisson algebra}
\newcommand{\pb}{Poisson bracket}
\newcommand{\la}{Lie algebra}
\newcommand{\lsa}{Lie subalgebra}

\newcommand{\ba}{basic algebra}
\newcommand{\fb}{{\mathfrak b}}
\newcommand{\fa}{{\mathfrak a}}
\newcommand{\fg}{{\mathfrak g}}

\newcommand{\fj}{{\mathfrak j}}
\newcommand{\ft}{{\mathfrak t}}
\newcommand{\T}{{T^*\!\,}}
\newcommand{\codim}{{\mbox{codim }}}

\def\endproof{\hfill $\Box$}

\def\bc{{\bf C}}
\def\r{{\bf R}}
\def\z{{\bf Z}}

\newtheorem{thm}{Theorem}
\newtheorem{lem}{Lemma}
\newtheorem{cor}[thm]{Corollary}
\newtheorem{prop}[thm]{Proposition}
\newtheorem{defn}{Definition}
\newtheorem{conj}{Conjecture}

\def\f #1,#2.{\textstyle{#1\over #2}}
\def\hlf{{\frac{1}{2}}}
\def\bc{{\bf C}}
\def\sp{{\rm span}}

\begin{document}


\title{Obstructions to Quantization\footnotetext{To appear in:
{\sl The Juan Simo Memorial Volume}, Marsden, J. \& Wiggins, S., Eds.
(Springer, New York) (1998).}}

\author{{\bf Mark J. Gotay} \\ 
\\  Department of Mathematics \\ University of Hawai`i \\ 2565 The
Mall \\ Honolulu, HI 96822  USA} 

\date{July 24, 1998 \\ (Revised September 9, 1998)}

\maketitle


\maketitle


\begin{abstract} Quantization is not a straightforward
proposition, as demonstrated by Groenewold's and Van~Hove's
discovery, more than fifty years ago, of an ``obstruction'' to
quantization. Their ``no-go theorems'' assert that it is in
principle  impossible to consistently quantize every classical
polynomial observable on the phase space ${\bf R}^{2n}$ in a
physically meaningful way. Similar obstructions have been recently
found for 
$S^2$ and $T^*\!S^1$, buttressing the common belief that no-go
theorems should hold in some generality. Surprisingly, this is not
so---it has just been proven that there are no obstructions to
quantizing either $T^2$ or $T^*\!\,\bf R_+$.

In this paper we work towards delineating the circumstances under
which such obstructions will appear, and understanding the mechanisms
which produce them. Our objectives are to conjecture---and in some
cases prove---generalized \gr-\vh\ theorems, and to determine the
maximal Lie subalgebras of observables which can be consistently
quantized. This requires a study of the structure of Poisson algebras
of symplectic manifolds and their representations. To these ends we
include an exposition of both prequantization (in an extended sense)
and quantization theory, here formulated in terms of ``basic algebras
of observables.'' We then review in detail the known results for
$\r^{2n}$, $S^2$, $T^*\!S^1$, $T^2$, and $T^*\!\,\bf R_+$, as well
as recent theoretical work. Our discussion is independent of any
particular method of quantization; we concentrate on the structural
aspects of quantization theory which are common to all Hilbert
space-based quantization techniques. 
\end{abstract}


\begin{section}{Introduction}

Quantization---the problem of  constructing the quantum formulation
of a system from its classical description---has always been one of
the great mysteries of mathematical physics. It is generally
acknowledged that quantization is an ill-defined procedure which
cannot be consistently applied to all classical systems. While there
is certainly no extant quantization procedure which works well in all
circumstances, this assertion nonetheless bears closer scrutiny.

Already from first principles one encounters difficulties. Given that
the  classical description of a system is an approximation to its
quantum description, obtained in a macroscopic limit (when
$\hbar\to 0$), one expects that some information is lost in the limit.
So quantization should somehow have to compensate for this. But how
can a given quantization procedure select, from amongst the myriad of
quantum theories all of which have the same classical limit, the
physically correct one? 

In view of this ambiguity it is not surprising that the many
quantization schemes which have been developed over the years---such
as the physicists' original ``canonical quantization''
\cite{di} (and its modern formulations, such as geometric quantization
\cite{ki,So70,Wo}), Weyl quantization \cite{Fo} (and its successor
deformation quantization
\cite{Bayen e.a.,ri2,ri3,ri4}), path integral quantization \cite{GJ},
and the group theoretic approach to quantization \cite{AA,Is}, to cite
just some---have shortcomings. Rather, is it amazing that they work as
well as they do!

But there are deeper, subtler problems, involving the Poisson algebras
of classical systems and their representations. In this context the
conventional wisdom is that it is impossible to ``fully'' quantize any
given classical system---regardless of the particular method
employed---in a way which is consistent with the physicists'
Schr\"odinger quantization of
${\bf R}^{2n}$. (We will make this somewhat nebulous statement precise
later.) In other words, the assertion is that there exists a universal
``obstruction'' which forces one to settle for something less than a
complete and consistent quantization of {\it any\/} system. Each
quantization procedure listed above evinces this defect in various
examples.

That there are problems in quantizing even simple systems was observed
very early on. One difficulty was to identify the analogue of the
multiplicative structure of the classical observables in the quantum
formalism. For instance, consider the quantization of ${\bf R}^{2n}$
with canonical coordinates
$\{q^i,p_i\,|\,i = 1,\dots,n\}$, representing the phase space of a
particle moving in $\r^n$. For simple observables the ``product
$\ra$ anti-commutator'' rule worked well. But for more complicated
observables (say, ones which are quartic polynomials in the positions
and momenta), this rule leads to inconsistencies. (See
\cite[\S 4]{a-b},
\cite[\S1.1]{Fo} and \S \S 5.1 and 6.5 for discussions of these
factor-ordering ambiguities.) Of course this, in and by itself, might
only indicate the necessity of coming up with some subtler
symmetrization rule. But attempts to construct a quantization map
also conflicted with Dirac's ``Poisson bracket $\ra$ commutator''
rule. This was implicitly acknowledged by Dirac
\cite[p.~87]{di}, where he made the now famous hedge:

\begin{quote}
\em ``The strong analogy between the quantum P.B.
$\ldots$ and the classical P.B.
$\ldots$ leads us to make the assumption that the quantum P.B.s, {\rm
or at any rate the simpler ones of them,} have the same values as the
corresponding classical P.B.s.''
\end{quote}

\noi In any case, as a practical matter, one was forced to limit the
quantization to relatively ``small'' Lie subalgebras of classical
observables which could be handled without ambiguity (e.g.,
polynomials which are at most quadratic in the $p$'s and the
$q$'s, or observables which are affine functions of the 
positions or of the momenta). 

Then, in 1946, Groenewold \cite{Gr} showed that the search for an
``acceptable'' quantization map was futile. The strong version of his
``no-go'' theorem states that one cannot consistently quantize the
Poisson algebra of all polynomials in the $q^i$ and $p_i$ on ${\bf
R}^{2n}$ as symmetric operators on some Hilbert space $\h,$ subject to
the requirement that the
$q^i$ and  $p_i$ be irreducibly represented.\footnote{\,There are
actually two variants of \gr's theorem (``strong'' and ``weak''); both
will be discussed in \S 5.1.} Van~Hove subsequently refined \gr's
result \cite{vH1}. Thus it is {\it in principle} impossible to
quantize---by {\it any\/} means---every classical observable on
${\bf R}^{2n}$, or even every polynomial observable, in a way
consistent with Schr\"odinger quantization (which, according to the
Stone-\vn\ theorem, is the import of the irreducibility requirement on
the $p$'s and $q$'s). At most, one can consistently quantize certain
Lie subalgebras of observables, for instance the ones mentioned in the
preceding paragraph. 


Of course, \gr's remarkable result is valid only for the classical
phase space $\r^{2n}.$ The immediate problem is to determine whether
similar obstructions appear when trying to quantize other symplectic
manifolds. Little was known in this regard, and only in the mid 1990's
have other examples come to light. A few years ago an obstruction was
found for $S^2$, representing the (internal) phase space of a massive
spinning particle \cite{GGH}. It was shown that one cannot
consistently quantize the Poisson algebra of spherical harmonics
(thought of as polynomials in the components
$S_i$ of the spin angular momentum vector $\bf S$), subject to the
requirement that the $S_i$ be irreducibly represented. This is a
direct analogue for $S^2$ of
\gr's theorem. Moreover, just recently it was shown that
the symplectic cylinder $T^*\!S^1$, which plays a role in geometric
optics, exhibits a similar obstruction \cite{GG1}. Combined with
the observations that $S^2$ is in a sense at the opposite extreme from
$\r^{2n}$ insofar as symplectic manifolds go, and that $T^*\!S^1$
lies somewhere in between, these results indicate that no-go theorems
can be expected to hold in some generality. But, interestingly enough,
they are {\em not\/} universal: It is possible to explicitly construct
a quantization of the full Poisson algebra of the torus $T^2$ in
which a suitable irreducibility requirement is imposed \cite{Go}. It
is also possible to quantize certain noncompact phase spaces, e.g.
$T^*\!\,\bf R_+$ \cite{GGra}. An important point, therefore, is to
understand the mechanisms which are responsible for these divergent
outcomes. 

Our goal here is to study obstructions to the
quantization of the Poisson algebra of a symplectic manifold. We will
review the known examples in some detail, and give a careful
presentation of prequantization (in an extended sense) and
quantization, with a view to conjecturing a generalized \gr-\vh\
theorem and in particular delineating the circumstances under which it
can be expected to hold. Already some results have been established
along these lines, to the effect that under certain circumstances
there are obstructions to quantizing both compact and noncompact
symplectic manifolds 
\cite{GGG,GGra,GG2,GM}. Despite these recent advances, many
interesting and difficult problems remain. Our discussion will be
independent of any particular method of quantization; we concentrate
on the structural aspects of quantization theory which are common to
all Hilbert space-based quantization techniques. 

The present paper is a revised and updated version of the review
article ``Obstruction Results in Quantization Theory,'' which was
published in 1996 in the {\it Journal 
of Nonlinear Science}
\cite{GGT}. Since a number of new results and examples have been
obtained since that article appeared, we thought it useful to
provide a more current summary of the field. As well, a
number of the concepts and constructions of that paper have evolved
over time, and we have amended the paper accordingly. We have also
taken this opportunity to correct a number of misprints and minor
errors.

We express our appreciation to V. Aldaya, C. Atkin, P. Chernoff, G.
Folland, V. Ginzburg, P. Jorgensen, J. Velhinho, and N. Wildberger,
for  many helpful conversations and for sharing insights, and to J.
Grabowski, G. Tuynman, and especially H. Grundling, for fruitful
collaborations. This research was supported in part by NSF grant DMS
96-23083.

\end{section}


\begin{section}{Prequantization}

Let
$(M,\omega)$ be a fixed $2n$-dimensional connected symplectic
manifold with associated Poisson algebra
$\big(C^\infty(M),\{\cdot\, ,\cdot\}\big)$, where
$\{\cdot\, ,\cdot\}$ is the Poisson bracket.  

To start the discussion, we state what it means to ``prequantize'' a
Lie algebra of observables.

\begin{defn}$\,\,$ {\rm Let $\oo$ be a Lie
subalgebra of $\p$ containing the constant function $1$. A {\it
prequantization\/} of
$\oo$ is a linear map
$\q$ from $\oo$ to the linear space Op($D$) of symmetric operators
which preserve a fixed dense domain $D$ in some separable Hilbert space
$\h$, such that for all $f,g \in \oo$,

\begin{description}
\item \rule{0mm}{0mm}
\begin{enumerate}
\vspace{-4.5ex}
\item[(Q1)] ${\cal Q}(\{f,g\}) =
\frac{i}{\hbar}[{\q}(f),{\q}(g)]$,
\vskip 6pt
\item[(Q2)]  ${\cal Q}(1) = I$, and
\vskip 6pt
\item[(Q3)] if the Hamiltonian vector field $X_f$ of $f$ is complete,
then
$\q(f)$ is essentially self-adjoint on $D$.
\end{enumerate}
\end{description}

\noi If $\oo = \p$, the prequantization is said to be {\it full\/}. 
 A prequantization $\q$ is {\it nontrivial\/} provided $\codim
\ker \q > 1$; otherwise $\q$ factors through a representation
of $\oo/\ker \q$ with $\dim (\oo/\ker \q) \leq 1.$}
\end{defn}

\noi {\it Remarks.} 1. By virtue of (Q1) a prequantization $\q$
of
$\oo$ is essentially a Lie representation of
$\oo$ by symmetric operators. (More precisely: If we set $\pi(f) =
\frac{i}{\hbar}\,\q(f)$, then $\pi$ is a true Lie 
representation by skew-symmetric operators on $D$ equipped with the
commutator bracket. We will blur the distinction between $\pi$ and
$\q$.) In this context there are several additional requirements we
could place upon
$\q$, such as irreducibility and integrability. However, we do not
want to be too selective at this point, so we do not insist on these;
they will be discussed as the occasion warrants.

2. Condition (Q2) reflects the fact that if an observable
$f$ is a constant $c$, then the probability of measuring $f = c$ is
one regardless of which quantum state the system is in. It also serves
to eliminate some ``trivial'' possibilities, such as the regular
representation $f
\mapsto -i\hbar X_f$ on
$L^2(M,\omega^n)$.

3. Regarding (Q3), we remark that in contradistinction with \vh\
\cite{vH1}, we do not confine our considerations to only those
classical observables whose Hamiltonian vector fields are complete.
Rather than taking the point of view that ``incomplete'' classical
observables cannot be quantized, we simply do not demand that the
corresponding quantum operators be essentially self-adjoint
(``e.s.a.''). We do not imply by this that symmetric operators which
are not e.s.a. are acceptable as physical observables; as is well
known, this is a controversial point.

4.  Notice that no assumptions are made at this stage regarding
the multiplicative structure on $\p$ vis-\`a-vis $\q$. This is partly
for historical reasons: In classical mechanics the Lie algebra
structure has played a more dominant role than the associative algebra
structure, so it is natural to concentrate on the former. This
is also the approach favored by Dirac \cite{di} and the geometric
quantization theorists \cite{So70,Wo}. For more algebraic treatments,
see
\cite{As,Em,vn}. The associative algebra structure is emphasized to a
much greater degree in deformation quantization theory
\cite{Bayen e.a.,ri2,ri3}. We shall make some comments on
this as we go along; see especially \S\S  5.1 and 6.5.

\ms

Prequantizations in this broad sense (even full ones) are usually easy
to construct, cf. \cite{Ch2,Ur,Wo}. Van Hove was the first to
construct a full prequantization of $\ci({\bf R}^{2n})$ \cite{vH1}. It
goes as follows: The Hilbert space $\h$ is
$L^2({\bf R}^{2n})$, the domain $D$ is the Schwartz space ${\cal
S}({\bf R}^{2n},\!\bc)$ of rapidly decreasing smooth
complex-valued functions (for instance), and for $f
\in
\ci({\bf R}^{2n})$,
\begin{equation}
\q(f) = -i\hbar\sum_{k=1}^n\left[\frac{\partial f}{\partial
p_k}\left(\frac{\partial}{\partial q^k} -\frac{i}{\hbar}\,p_k\right) -
\frac{\partial f}{\partial q^k}\frac{\partial}{\partial p_k}\right] +
f.
\label{eq:preq}
\end{equation}

As luck would have it, however, prequantization representations tend
to be flawed. For example, the Van Hove prequantization of $\ci({\bf
R}^{2n})$, when restricted to the Heisenberg subalgebra h$(2n) \cong $
span$\{1,p_i,q^i\,|
\,i=1,\ldots,n\}$, is not unitarily equivalent to the Schr\"odinger
representation (which it ought to be, in the context of a particle
moving in
$\r^n$ with no superselection rules)
\cite{bl1,Ch1}. (Recall that the {\it Schr{\"o}dinger
representation\/} of h$(2n)$ is defined by\footnote{\,We denote
multiplication operators as functions.}
\begin{equation} q^i \mapsto q^i,\;\;\; p_j \mapsto -i\hbar \,
\partial /{\partial q^j},\;\;\; \mbox{and}\;\;\; 1 \mapsto I
\label{eq:srep}
\end{equation} 

\noi on the domain $\s (\r^n,\!\bc) \subset
L^2({\r^{n}}).$  It is irreducible in the sense given in the next
section.) There are various ways to see this; we give
\vh's original proof
\cite[\S 17]{vH1} as it will be useful later. Take $n=1$ for
simplicity. First, define a unitary operator $F$ on $L^2(\r^2)$ by
\[(F\psi)(p,q) = \frac{1}{\sqrt h}\int^{\infty}_{-\infty}
e^{ipv/\hbar}\psi(v,q-v)\,dv.\] 

\noi Then for each fixed $j = 0,1,\ldots$ take
$\h_j$ to be the closure in $L^2(\r^2)$ of the linear span of elements
of the form
$Fh_{jk}$, where $h_{jk}(p,q) = h_j(p)h_k(q)$ and 
\begin{equation}
\label{eq:hermite}
h_k(q) = e^{q^2/2} \frac{d^k}{dq^k} e^{-q^2}
\end{equation}

\noi is the Hermite
function of degree $k$. Now from (\ref{eq:preq}), 
\[\q(q) = i\hbar \, \frac{\partial}{\partial p} + q,\;\;\;\q(p) =
-i\hbar \,\frac{\partial}{\partial q}.\] 

\noi These operators are e.s.a.\   on
$\s (\r^2,\!\bc),$ and one may verify that they strongly com\-mute with
the orthogonal projectors onto the closed subspaces
$\h_j$.\footnote{\,Recall that two e.s.a.\ (or, more generally,
normal) operators {\it strongly commute\/} iff their spectral
resolutions commute, cf.
\cite[\S VIII.5]{ReSi}. Two operators $A,\,B$ {\it weakly commute\/}
on a domain $D$ if they commute in the ordinary sense, i.e., $[A,B]$
is defined on $D$ and vanishes.} Thus the
\vh\ prequantization of $\ci(\r^2)$ is reducible when restricted to
the Heisenberg subalgebra and hence does not produce the Schr\"odinger
representation. Moreover the association $Fh_{jk}(p,q) \mapsto
c_{j}h_k(q),$ where the $c_{j}$ are normalization constants, provides
a unitary equivalence of each subrepresentation of h(2) on
$\h_j$ with the Schr\"odinger representation on $L^2(\r)$,  from which
we see that the multiplicity of the latter is infinite in the
\vh\ representation. The \vh\ representation suffers from other
defects as well \cite[\S4.5.B]{Z}.

Likewise, the Kostant-Souriau prequantizations of
$S^2$ do not reproduce the familiar spin representations of the
special unitary algebra su(2). We realize $S^2$ as a coadjoint orbit
in $\mbox{su(2)}^*$ according to ${\bf S\cdot S} = s^2$, where ${\bf
S} = (S_1,S_2,S_3)$ is the spin vector and
$s > 0$ is the classical spin. It comes equipped with the symplectic
form
\begin{equation}
\omega=\frac{1}{2s^2}\sum_{i,j,k = 1}^3\epsilon_{ijk}\,S_i\,
dS_j\wedge dS_k. 
\label{eq:sfs2}
\end{equation}

\noi Now the de Rham class $[\!\,\omega/h\!\,]$ is integral iff $s =
\frac{n}{2}\hbar$, where
$n$ is a positive integer, and the corresponding Kostant-Souriau
prequantum line bundles can be shown to be
$L^{\otimes n}$ where $L$ is the dual of the universal line bundle over
$S^2$ \cite{ki}. The corresponding prequantum Hilbert spaces $\h_n$
can thus be identified with spaces of square integrable sections
$\psi$ of these bundles w.r.t. the inner product 
\[\langle \psi,\phi \rangle = \frac{i}{2\pi}
\int_{\scriptstyle {\bf C}}\frac{\overline{\psi(z)}\phi(z)\,dz
\wedge d{\bar z}}{(1+z{\bar z})^{n+2}}\]

\noi where $z = (S_1 + iS_2)/(s - S_3)$, cf. \cite{Wo}. But these
$\h_n$ are infinite-dimensional, whereas the standard representation
spaces for quantum spin $s = \frac{n}{2}\hbar$ have dimension $n+1$.

In both examples the prequantization Hilbert spaces are ``too big.''
The main problem is how to remedy this, in other words, how to modify
the notion of a prequantization so as to yield a genuine
{\it quantization.}

It is here that the ideas start to diverge, because there is less
agreement in the literature as to what constitutes a quantization map.
Some versions define it as a prequantization, not necessarily defined
on the whole of
$\p$, which is irreducible on a ``basic set''
$\fb\subset\p$ \cite{ki}. This is in line with the group theoretical
approach to quantization \cite{Is}, in which context $\fb$ is
realized as the Lie algebra of a symmetry group;\footnote{\,We
typically identify an abstract Lie algebra with its isomorph in
$\p$.} quantization should then  yield an irreducible representation
of this algebra.  For example, when
$M=\r^{2n}$, one usually takes $\fb$ to be the Heisenberg algebra
${\mbox h(2}n) \cong {\sp}\{1, p_i, q^i\,|\,i=1,\ldots, n\}$ of
polynomials of degree at most one. Similarly, when
$M=S^2$, one takes for $\fb$ the special unitary algebra ${\rm
su(2)} \cong {\sp}\{S_1, S_2, S_3\}$ of spherical harmonics of
degree one. We will plumb in detail the rationale behind these
choices of $\fb$ in the next section.

A different approach to quantization is to require a prequantization
$\q$ to satisfy some ``Von~Neumann rule,'' that is, some given
relation between the classical multiplicative structure of $\p$ and
operator multiplication on
$\h$. (Note that thus far in our discussion the multiplication on
$\p$ has been ignored, and it is reasonable to require that
quantization preserve at least some of the associative algebra
structure of $\p$, given that the Leibniz rule intertwines pointwise
multiplication with the Poisson bracket.) There are many different
types of such rules
\cite{Co,Fo,KLZ,KS,Ku,MC,vn}, the simplest being of the form 
\begin{equation}
\q(\varphi\circ f)=\varphi\big(\q(f)\big)
\label{eq:vnr}
\end{equation}

\noi for some distinguished observables
$f\in\p$, and certain smooth functions
$\varphi\in C^\infty(\r)$. (Technically, if $\varphi$ is not a
polynomial, then
$\q(f)$ must be e.s.a.\  for $\varphi\big(\q(f)\big)$ to be defined.)
In the case of
$M=\r^{2n}$, \vn\ states that the physical interpretation of the
quantum theory requires (\ref{eq:vnr}) to hold for {\em all\/} $f \in
\p$ and $\varphi\in C^\infty(\r)$
\cite{vn}. However, it is easy to see that this is impossible (simple
demonstrations are given in 
\cite{a-b,Fo} as well as \S 5.1 following); hence the qualifiers in
the definition above. In this example, one typically ends up imposing
the squaring
\vn\ rule
$\varphi(x) = x^2$ on elements of h(2$n$). The relevant rule for
the sphere turns out to be less intuitive; it takes the form
$\q(S_i\,\!^2) = a\q(S_i)^2 + cI$ for $i = 1,\,2,\,3$, where
$a,c$ are undetermined (representation-dependent) constants subject
only to the constraint that $a^2 + c^2 \neq 0$. Derivations of these
rules in these two examples are given in \S5 and \cite{GGH}. 

Another type of quantization is obtained by ``polarizing'' a
prequantization representation
\cite{Wo}. Following Blattner \cite{bl1}, we paraphrase it
algebraically as follows. Start with a {\it polarization\/}, i.e., a
maximally commuting Lie subalgebra $\A$ of $\p$. Then require for
the quantization map
$\q$ that the image $\q(\A)$ be ``maximally commuting'' as operators.
(If
$\q(\A)$ consists of bounded operators, this means that the weak
operator closure of the *-algebra generated by $\q(\A)$ ($=
\q(\A)''$) is maximally commuting in $B(\h )$. If
$\q(\A)$ contains unbounded operators, one should look for a
generating set of normal operators in $\q(\A)$, and require that the
Von~Neumann algebra generated by their spectral projections is
maximally commuting.) One can then realize the Hilbert space
$\h $ as an $L^2$-space over the spectrum of this \vn\ algebra on
which this algebra acts as multiplication operators. There will also
be a cyclic and separating vector for such an algebra, which provides
a suitable candidate for a vacuum vector. Thus another motivation for
polarizations is that a maximally commuting set of observables
provides a set of compatible measurements, which can determine the
state of a system. When $M=\r^{2n}$, one often takes the ``vertical''
polarization $\A =
\big\{f(q^1,\ldots,q^n)\big\}$, in which case one recovers the usual
position or coordinate representation. However, in some instances,
such as $S^2$, it is useful to broaden the notion of polarization to
that of a maximally commuting subalgebra of the {\em
complex\-i\-fi\-ed\/} Poisson algebra
$\ci (M,\!\bc).$ Then, thinking of $S^2$ as $\bc
P^1$, we may take the ``antiholomorphic'' polarization $\A =
\{f(z)\},$ which leads to the usual representations for spin. For
treatments of polarizations in the context of deformation
quantization, see
\cite{Fr,He}.

Thus, informally, a ``quantization'' could be defined as a
prequantization which incorporates one (or more) of the three
additional requirements above (or possibly even others). Before
proceeding, however, there are two points we would like to make.

The first is that it is of course not enough to simply state the
requirements that a quantization map should satisfy; one must also
devise methods for implementing them in examples. Thus geometric
quantization theory, for instance, provides a specific technique for
polarizing certain (Kostant-Souriau) prequantization representations
\cite{bl1,ki,So70,Wo}. However, as we are interested here in the
structural aspects of quantization theory, and not in specific
quantization schemes, we do not attempt to find such implementations.

Second, these three approaches to a quantization map are not
independent; there exist subtle connections between them which are not
well understood. For instance, demanding that a prequantization be
irreducible on some basic algebra typically leads to the appearance of
\vn\ rules; this is how the \vn\ rules for $\r^{2n}$ and $S^2$
mentioned above arise. We will delineate these connections in specific
cases in
\S 5, and more generally in \S 7.

\ms

At the core of each of the approaches above is the imposition---in
some guise---of an irreducibility requirement, which is used to ``cut
down'' a prequantization representation. Since this is most apparent
in the first approach, we will henceforth concentrate on it. We will
tie in the two remaining approaches as we go along.

So let $\oo$ be a Lie subalgebra of $\p$, and suppose that $\fb
\subset \oo$ is a ``basic algebra'' of observables. Provisionally, we
take a {\it quantization} of the pair $({\oo},{\fb})$ to mean a
prequantization $\q$ of $\oo$ which (among other things) irreducibly
represents $\fb$. In the next section we will make this more precise,
as well as examine in detail the criteria that $\fb$ should satisfy.

Natural issues to address for quantizations are existence, uniqueness
and classification, and functoriality. For {\em pre\/}quantizations
these questions already have partial answers from geometric
quantization theory. So for instance we know that if $(M,\omega)$
satisfies the integrality condition
$[\!\,\omega/h\,\!]\in H^2(M,\z)$, then full prequantizations of the
Poisson algebra $\p$ exist, and that certain types of these---the
Kostant-Souriau prequantizations---can be  classified
cohomologically  \cite{Ur,Wo}. For some limited types of manifolds the
functorial properties of these prequantizations were considered by
Blattner
\cite{bl1}. However, as there are prequantizations not of the
Kostant-Souriau type \cite{av,Ch2}, these questions are still open in
general (especially for manifolds which violate the integrality
condition
\cite{We}). 

For quantization maps these questions are far more problematic. Our
main focus will be on the existence of both {\it full
quantizations}, by which we mean a
quantization of $(\p,\fb)$ for some appropriately chosen basic
algebra $\fb$, and {\it polynomial quantizations}, by which we mean a
quantization of 
$(P(\fb),\fb)$,
where $P(\fb)$ is the Poisson algebra of polynomials generated by
$\fb$. As indicated earlier, these are not completely understood in
general, although substantial progress has been made in the past
several years. In our terminology, the classical (strong) result of
Groenewold states that there is no quantization of
$\big(P({\rm h}(2n)),{\rm h}(2n)\big)$ on $\r^{2n}$, while the more
recent results of \cite{GGH} and \cite{GG1} imply the same for
$\big(P({\rm su}(2)),{\rm
su}(2)\big)$ on $S^2$
and $\big(P({\rm e}(2)),{\rm
e}(2)\big)$ on $T^*\!S^1$, respectively, where e(2) is the
Euclidean algebra (cf. \S 5.3).  On the other hand, nontrivial
polynomial quantizations do exist: One can construct such a 
quantization of
$T^*\!\,{\bf R}_+$ with the affine algebra a(1)
\cite{GGra}. In fact, full quantizations exist as well; there is one
of $T^2$ with $\fb$ the Lie algebra of trigonometric
polynomials of mean zero \cite{Go}. However, it does seem that
nonexistence results are the rule. In the absence of a full (resp. a 
polynomial) quantization, then, it is important to determine the
maximal Lie subalgebras
$\oo$ of $\p$ (resp. $P(\fb)$) for which $(\oo,\fb)$ can be quantized.
This we will investigate for
$\r^{2n}$, $S^2$, and $T^*\!S^1$ in \S 5. At present,
questions of uniqueness and classification can only be answered in
specific examples.

\end{section}


\begin{section}{Basic Algebras of Observables}

Our first goal here is to make clear what we mean by a \ba\ of
observables
$\fb\subset\p$. Such algebras, in one way or another, play an
important role in many quantization methods, such as geometric
quantization
\cite{ki}, deformation quantization \cite{Bayen e.a.,Fr} and also the
group theoretic approach \cite{Is}.

We start with the most straightforward case, that of an ``elementary
system'' in the terminology of Souriau
\cite{So70,Wo}. This means that $M$ is a homogeneous space for a
Hamiltonian action of a finite-dimensional Lie group $G$. The appeal
of an elementary system is that it is a classical version of an
irreducible representation: Using the transitive action of
$G$, one can obtain any classical state from any other one, in direct
analogy with the fact that every nonzero vector in a Hilbert space
$\h$ is cyclic for an irreducible unitary representation (``IUR'') of
$G$ on
$\h$ \cite[\S 5.4]{b-r}. Now notice that the span
$\fj$ of the components of the associated (equivariant) momentum
map satisfies:
\begin{description}
\item \rule{0mm}{0mm}
\begin{enumerate}
\vspace{-4.5ex}
\item[(J1)] $\fj$ is a finite-dimensional Lie subalgebra of $\p$,
\vskip 6pt
\item[(J2)] the Hamiltonian vector fields
$X_f,{f\in\fj}$, are complete, and
\vskip 6pt
\item[(J3)]  $\{X_f \,|\,{f\in\fj}\,\}$ spans $TM$.
\end{enumerate}
\end{description}

\noi  For both $M=\r^{2n}$ and $S^2$, the basic algebras are precisely
of this type: {}From the elementary systems of the Heisenberg group
H(2$n$) acting on
$\r^{2n}$, and the special unitary group SU(2) acting on $S^2$, we
have for
$\fj$ the spaces span$\{1,p_i,q^i\,|\,i = 1,\ldots,n\}$ and
span$\{S_1,S_2,S_3\}$, respectively. The same is true for $M = T^*\!
S^1$ and $\T \r_+$, as explained in \S 5.

Property (J3) is just an infinitesimal restatement of transitivity,
and so we call a subset of $C^{\infty}(M)$ {\it transitive\/} if
it satisfies this condition. Kirillov \cite{ki} uses the terminology
``complete,'' motivated by the fact that such a set of observables
locally separates classical states. (If a set of observables {\em
globally\/} separates classical states, we call it {\em
separating.}) In this regard, the finite-dimensionality criterion in
(J1) plays an important role operationally: It guarantees that a
finite number of measurements using this collection of observables
will suffice to distinguish any two nearby states.

A Lie subalgebra $\fb \subset \p$  satisfying (J1)--(J3) is a
prototypic basic algebra. However, there need not exist basic algebras
in this sense for arbitrary $M$. For instance, if
$M = T^2$, the self-action of the torus is not Hamiltonian, so there
is no momentum map. Thinking of $T^2$ as
$\r^{2}/\z^{2}$, a natural choice of \ba\ is then the Lie algebra
$\mathfrak t$ generated by
\[{\cal T} = \{\sin 2\pi x,\cos 2\pi x,\sin 2\pi y,\cos 2\pi y\}.\]

\noi This \la\ (viz. the set of trigonometric polynomials of
mean zero) is \id. While perhaps unpleasant, this is in fact
unavoidable: It follows from Proposition \ref{prop:char} below that
there is no \fd\ basic algebra on $T^2$. However, in keeping with the
discussion above, note that $\mathfrak t$ is finitely generated,  
and one can
use this generating set to separate states.

We will therefore dispense with the finite-dimensionality
assumption, and instead merely require that $\fb$ be finitely
generated. One then still has a finite number of ``basic observables''
with which to distinguish states. Thus we make:

\begin{defn} {\rm \, A {\it basic algebra of observables\/} $\fb$ is a
\lsa\ of $\p$ such that:\footnote{\,This definition differs from that
given in \cite{GGT} in three regards: It is phrased in terms of basic
{\it algebras} as opposed to basic {\it sets}, we no longer
insist that $1 \in \fb$ (this is superfluous), and we have
strengthened (B3) by requiring that $\fb$ be separating.}
\begin{description}
\item \rule{0mm}{0mm}
\begin{enumerate}
\vspace{-5ex}
\item[(B1)] $\fb$ is finitely generated,
\vskip 6pt
\item[(B2)] the Hamiltonian vector fields
$X_f,{f\in\fb}$, are complete, 
\vskip 6pt
\item[(B3)] $\fb$ is transitive and separating, and
\vskip 6pt
\item [(B4)] $\fb$ is a minimal \la\ satisfying these requirements.
\end{enumerate} \end{description}}
\end{defn}

We spend some time elaborating on this definition. First, the
completeness condition (B2) guarantees that a basic observable
generates a one-parameter group of canonical transformations. In view
of (Q3), it is the classical analogue of the requirement that an
operator representing a physically observable quantity should be
e.s.a., whence it generates a  one-parameter group of unitary
transformations. 

Next consider the transitivity requirement in (B3). When $\fb$ is
finite-dimen\-sional, it together with (B2) enables us to integrate
$\fb$ to a transitive group action on 
$M$. Indeed, the map $f \mapsto X_f$ can be thought of as an
action of $\fb$ on $M$. By (B2) the
vector fields $X_f$ are complete and so by a theorem of Palais
\cite[Thm.~2.16.13]{V} this action of $\fb$ can be
integrated to an action of the corresponding simply connected Lie
group $B$. Condition (B3) implies that this
action is locally transitive and thus globally transitive as $M$ is
connected.

As part of (B3) we also require that $\fb$ globally separate
classical states. This ensures that $\fb$ accurately reflects the
topology of $M$ \cite{Ve}. Without it, e.g., the \la\ $\mathfrak t$
defined above could equally well live on either $\r^2$ or $T^2$ (or
even ``halfway between,'' on $\T\! S^1$); measurements using
$\mathfrak t$ could not distinguish amongst these phase spaces.

Although a transitive set of observables is locally separating, it
need not be (globally) separating. Conversely, a separating set of
observables need not be everywhere transitive. So these two
conditions are distinct.

While
(B3) is geometrically natural, there are other conditions one might
use in place of it. By way of motivation, consider a unitary
representation
$U$ of a Lie group
$G$ on a Hilbert space $\h$. The representation
$U$ is irreducible iff the *-algebra ${\cal U}$ of bounded operators
generated by $\{U(g)\,|
\,g \in G\}$ is irreducible, in which case we have the following
equivalent characterizations of irreducibility:

\begin{description}
\item \rule{0mm}{0mm}
\begin{enumerate}
\vspace{-4.5ex}
\item[(I1)] The commutant ${\cal U}'= \bc I,$ and
\vskip 6pt
\item[(I2)] the weak operator closure of $\cal U$ is the algebra of
all bounded operators: $\overline{\cal U}^w=B(\h)\quad(={\cal U}'').$

\end{enumerate}\end{description}

\noi That (I1) is equivalent to irreducibility is the content of
Schur's Lemma. Property (I2) means that all bounded operators can be
built from those in
$\cal U$ by weak operator limits. It follows from (I1), the \vn\
density theorem \cite[Cor.~2.4.15]{BR}, and the fact that
${\cal U}'=\big(\,\overline{\cal U}^w\big)'$. Clearly (I2) implies
(I1). 

These restatements of irreducibility have the following classical
analogues for a set ${\cal F}$ of observables:

\begin{description}
\item \rule{0mm}{0mm}
\begin{enumerate}
\vspace{-4.5ex}
\item[(C1)] $\{f,g\}=0$ for all
$f\in {\cal F}$ implies
$g$ is constant, and
\vskip 6pt
\item[(C2)] the Poisson algebra of polynomials generated by ${\cal F}$
forms a dense subspace in
$C^{\infty}(M)$.

\end{enumerate}\end{description}

\noi For (C2) a topology on $C^{\infty}(M)$ must be decided on, and we
will use the topology of uniform convergence on compacta of a function
as well as its derivatives. 

Because the algebraic structures of classical and quantum mechanics
are different,  (C1) and (C2) lead to inequivalent notions of
``classical irreducibility.'' It is not difficult to verify that (C1)
$\Leftarrow$ (B3) $\Leftarrow$ (C2) strictly. In principle either of
(C1) or (C2) could serve in place of (B3). Indeed, since on
$C^{\infty}(M)$ one has two algebraic operations, it is natural to
consider irreducibility in either context: in terms of the
multiplicative structure (C2), or the Poisson bracket (C1). However,
it turns out that (C1) is too weak for our purposes, while (C2) is too
strong.

The nondegeneracy condition (C1) is equivalent to the statement that
observables in $\fb$ locally separate states almost everywhere
\cite{ki}. It is also equivalent to the statement that the Hamiltonian
vector fields of elements of $\fb$ span the tangent spaces to
$M$ almost everywhere. 
Consequently, it will not do to replace (B3) by (C1) in the
definition of basic algebra, for
then as shown below the \la\ $\ft$ on $T^2$ would no longer be
minimal, which seems both awkward and unreasonable. Furthermore,
unlike (B3), (C1) has the defect that the simply connected covering
group of
$\fb$ need not act transitively on
$M$. This happens for the symplectic algebra sp$(2,\r) \cong$
span$\{p^2,pq,q^2\}$ on
$\r^{2}$. Condition (C2)  
fails for the affine algebra a(1) $\cong \sp\{pq,q^2\}$ on
$\T \r_+$ since, e.g., $\ci (\T \r_+)$ contains functions which blow
up as
$q \ra 0$ along with all their $q$-derivatives, and such
functions cannot be approximated by polynomials in the elements of
a(1). On the other hand, all these examples satisfy (B3),
which shows that this is a reasonable condition to impose.


Finally, the minimality condition (B4) is crucial. {}From a physical
or operational point of view, it is not obvious that it is necessary,
as long as
$\fb$ is finitely generated. But the quantization of a pair
$(\oo,\fb)$ with $\fb$ nonminimal in this sense can lead to
physically incorrect results. 

Here is an illustration. First observe that the extended
symplectic group HSp(2$n$,$\r$) (which is the semidirect product of
the symplectic group Sp(2$n$,$\r$) with the Heisenberg group H(2$n$))
acts transitively on
$\r^{2n}$. This action has a momentum map whose components consist of
all inhomogeneous quadratic polynomials in the $q^i$ and
$p_i$. The corresponding Lie subalgebra $\fj \cong \mbox{hsp}(2n,\r)$
satisfies all the requirements for a \ba\ save minimality, since
h(2$n$) is a separating transitive subalgebra of hsp(2$n$,$\r$). Now
consider again the Van Hove prequantization
$\q$ of
$C^\infty(\r^{2n})$ for $n=1$. In
\cite[\S 17]{vH1} it is shown that $\q$ is completely reducible when
restricted to $\fj$. In fact, there exist exactly two nontrivial
HSp(2,$\r$)-invariant closed subspaces $\h_{\pm}$ in $L^2(\r^{2})$,
namely 
\[\h_+ = \bigoplus_{j \;{\rm even}}\h_j\;\;\;{\rm and}\;\;\; \h_- =
\bigoplus_{j
\;{\rm odd}}\h_j,\] 

\noi cf. \S 2. If we denote the corresponding subrepresentations of
$\fj$ on
${\cal S}(\r^{2},\!\bc) \cap \h_{\pm}$ by
$\q_{\pm}$, then it follows that $\q_{\pm}$ are quantizations of the
pair $(\fj,\fj).$ But these quantizations are physically unacceptable,
since---just like the full prequantization $\q$---they are reducible
when further restricted to h(2). On the one hand, asking for a
quantization of
$(\fj,\fj)$ in this context is clearly the wrong thing to do, since
compatibility with Schr\"odinger quantization devolves upon the
irreducibility of an h(2) algebra, not an hsp(2$,\r$) one. But
on the other hand, this example does make our point.

To illustrate the appropriateness of this definition,
consider again the torus and let
$\ft_k$ be the
\la s generated by the sets
\[{\cal T}_k = \{\sin 2\pi kx,\,\cos 2\pi kx,\,\sin 2\pi
ky,\,\cos 2\pi ky\}\]

\noi for $k = 1,2,\ldots$ Each
$\ft_k$ is transitive. But without the separation axiom
in (B3), none of the $\ft_k$ would be
minimal, since each contains the infinite descending series
$\ft_k \supset \ft_{2k}
\supset \cdots$. However, only
$\ft = \ft_1$ is separating, and in fact it is a minimal
\emph{separating} transitive subalgebra.

Other properties that basic algebras might be required to satisfy are
discussed in \cite{Is}. For our purposes, (B1)-(B4) will suffice.


\ms

It appears difficult to characterize \ba s on general symplectic
manifolds. In the compact case, however, we can be quite precise.
\begin{prop}
\label{prop:bs} 
Let $\fb$ be a \fd\ \ba\ on a compact symplectic manifold.
Then $\fb$ is compact and semisimple. In
particular, its center must be zero. 
\end{prop}

\noi \emph{Proof}. Define an inner product on $\fb$ according to
\begin{equation*}
\langle f,g \rangle = \int_M fg\,\,\omega^n.
\label{eq:ip}
\end{equation*}

\noi Using the identity
\begin{equation}
\label{eq:identity}
\{f,g\}\,\omega^n = n\, d(f\,dg \wedge \omega^{n-1})
\end{equation}

\noi together with Stokes' Theorem, we immediately verify that
\[\langle \{f,g\},h \rangle + \langle g,\{f,h\} \rangle = 0\]

\noi whence $\fb$ is compact \cite[\S1.2.6]{On}. 
As a consequence,
$\fb$ splits as the Lie algebra direct sum $\mathfrak z \oplus
\mathfrak s$, where
$\mathfrak z$ is the center of
$\fb$ and $\mathfrak s$ is semisimple \cite[Prop. 1.2.8]{On}. 

Now transitivity implies that any function which
Poisson commutes with every element of $\fb$ must be a constant, so
that $\mathfrak z \subseteq \r.$ But if equality holds then
$\mathfrak s$ would be a separating transitive subalgebra, thereby
violating (B4). Thus
$\mathfrak z = \{0\}$ and $\fb$ is semisimple.
\endproof

\ms

In particular, the proof shows that any reductive (and consequently
any compact) \ba\ must be semisimple.

There is no guarantee that a given symplectic manifold
will carry a basic algebra. Indeed, the next proposition shows that
those phase spaces which admit
\ba s form a quite restricted class.
\begin{prop} If a  connected symplectic manifold
$M$ admits a finite-di\-men\-sion\-al
\ba\ $\fb$, then $M$ is a coadjoint orbit in $\fb^*.$
In particular, when $M$ is compact it must be simply connected.
\label{prop:char}
\end{prop}

\noi \emph{Proof.} For if $\fb$ is a
\fd\ \ba\ on $M$ then, by our consider\-ations above, $M$ must be a
homogeneous Hamiltonian $B$-space, where $B$ is the simply connected
covering group of $\fb$ and the momentum map $J$ is given
by $\langle J(m),b \rangle = b(m)$ for $b \in \fb$ and $m \in M.$ 
The Kirillov-Kostant-Souriau Coadjoint Orbit Covering Theorem
\cite[Thm.~14.6.5]{m-r} then implies that $J: M \ra \fb ^*$ is a
symplectic local diffeomorphism of $M$ onto a coadjoint orbit $O
\subset \fb^*.$ Since by (B3) $\fb$ is separating, it follows that
$J$ is injective (for otherwise elements of $\fb$ cannot separate
points in $J^{-1}(\mu)$ for $\mu \in O$.) Thus $M$ is
symplectomorphic to $O$.

If $M$ is compact, then by Proposition~\ref{prop:bs} $\fb$ is
compact and semisimple. We conclude that $B$ is compact
\cite[p.~29]{On}. But the coadjoint orbits of a compact connected Lie
group are simply connected \cite[Thm.~2.3.7]{Fi}.
\endproof  

\ms
 
As $M$ is a homogeneous space for $B$, the last paragraph of
this proof shows that $M$ is compact iff
$\fb$ is compact iff $B$ is compact. 

Thus the symplectic algebra sp(2,$\r$) is not a \ba\ on $\r^2
\setminus \{\bf 0\}$, since the latter is not a coadjoint orbit.
(Note that $\sp\{p^2,pq,q^2\}$ satisfies all the criteria for a
\ba\ save the separation axiom.) Even if $M \subset \fg^*$ is a
coadjoint orbit,
$\fg$ need
\emph{not} form a \ba. An example is provided by $S^2$, which is a
coadjoint orbit in
${\rm su}(2)^* \oplus \{0\} \subset {\rm su}(2)^* \oplus {\rm
su}(2)^* \cong {\rm o}(4)^*$ with basic algebra ${\rm su}(2)$, not
o(4). In the compact case we can be more explicit:
\begin{prop}
Let $M$ be a maximal coadjoint orbit in $\fg^*$,
where $\fg$ is a compact semi\-simple \la. Then $M$
admits $\fg$ as a basic algebra.
\label{prop:co}
\end{prop}

The proof is given in \cite{GGG}.

Despite all this, $M$ may still carry \emph{infinite}-dimen\-sion\-al
\ba s, as happens for $T^2.$ Not much is known regarding these, and we
refer the reader to
\cite{Is} for further discussion (cf. especially \S 4.8.4).

\ms

We denote by $P(\fb)$ the polynomial algebra generated by
$\fb$. Since $\fb$ is a Lie algebra, $P(\fb)$ is a Poisson algebra.
Note that (\emph{i}) $P(\fb)$ is not necessarily free as an
associative algebra (cf. the examples in
\S5), and (\emph{ii}) by definition $\r \subset P(\fb)$. When
$P(\fb)$ is free, it can be identified with the symmetric algebra
$S(\fb)$ generated by $\fb$, but otherwise $P(\fb)$
is realized as the quotient of $S(\fb)$ by the associative ideal
generated by elements of the form $C - c$, where $C$ is a
``Casimir'' and $c$ is some constant (depending upon $C$ and
$M$).\footnote{\,A \emph{Casimir} is an element of the Lie center of
$S(\fb\,)$ which has no constant term.}
Note that $S(\fb)$ is itself a unital \pa , and that the canonical
projection is a \pa\ homomorphism. 
In general we will not distinguish
between
$P(\fb)$ and $S(\fb)$, and in examples where Casimirs are present we
will often work with representatives, i.e., on $S(\fb)$, without
explicitly stating so. Let
$P^k(\fb)$ denote the subspace of polynomials of minimal
degree at most $k$. (Since $P(\fb)$ is not necessarily free, the
notion of ``degree'' may not be well-defined, but that of ``minimal
degree'' is.) In the cases when degree does make sense, we let
$P_k(\fb)$ denote the subspace of homogeneous polynomials of degree
$k$, so that $P^k(\fb) = \oplus_{l = 0}^k P_l(\fb)$ (vector space
direct sum). We then also introduce $P_{(k)}(\fb) = +_{l \geq k}
P_l(\fb).$ Notice that
$P^1(\fb) = \fb$ or $\r \oplus \fb$, depending upon whether $1
\in \fb$ or not. When
$\fb$ is fixed in context, we simply write $P = P(\fb)$, etc.

\end{section}


\begin{section}{Quantization}

We are now ready to discuss what we mean by a ``quantization.'' Let
$\oo$ be a Lie subalgebra of $\p$, and suppose that $\fb \subset
\oo$ is a \ba\ of observables. Two eminently reasonable
requirements to place upon a quantization are irreducibility and
integrability \cite{b-r,fl,Is,ki}.

Irreducibility is of course one of the pillars of the quantum theory,
and we have already seen the necessity of requiring that quantization
represent
$\fb$ irreducibly. We must however be careful to give a precise
definition since the operators
$\q(f)$ are in general unbounded (although, according to (B2) and
(Q3), all elements of
$\q(\fb)$ are e.s.a.). So let
${\cal X}$ be a set of e.s.a.\  operators defined on a common
invariant dense domain $D$ in a Hilbert space $\h$. Then 
${\cal X}$ is {\it irreducible\/} provided the only bounded
self-adjoint operators which strongly commute with all $X \in {\cal
X}$ are multiples of the identity. While this definition is fairly
standard, and well suited to our needs, we note that other notions of
irreducibility can be found in the literature \cite{b-r,mmsv}; see
also \S 6.5. 

Given such a set ${\cal X}$ of operators, let $\uu ({\cal X})$ be the
$^*$-algebra generated by the unitary operators $\big\{\exp(it \ol X)
\,|\, t \in \r,\; X \in {\cal X}$\big\}, where $\ol X$ is the closure
of
$X$. Then by Schur's Lemma ${\cal X}$ is irreducible iff the only
closed subspaces of
$\h$ which are invariant under $\uu ({\cal X})$ are $\{0\}$ and $\h$.

Turning now to integrability, we first consider the case when \ba\ is
finite-dimensional. Then it is natural to demand that the Lie algebra
representation
$\q(\fb)$ on $D
\subset \h$ be {\it integrable\/} in the following sense: There
exists a unitary representation ${\Pi}$ of some Lie group with
Lie algebra $\fb$ on $\h$ such that
$\q(f) = -i\hbar \,d{\Pi}(f)\restriction D$ for all $f \in \fb$, where
$d{\Pi}$ is the derived representation of ${\Pi}.$ For this it is
{\em not\/} sufficient that elements of
$\fb$ quantize to e.s.a. operators on $D$ \cite[\S VIII.5]{ReSi}. But
integrability will follow from the following result of Flato et al.,
cf. \cite{fl} and
\cite[Ch.~11]{b-r}.
\begin{prop} Let ${\fg}$ be a real finite-dimensional Lie algebra,
and let $\pi$ be a representation of ${\fg}$ by skew-symmetric
operators on a common dense invariant domain $D$ in a Hilbert space
$\h$. Suppose that $\{\xi_1,\ldots,\xi_k\}$ generates ${\fg}$ by
linear combinations and repeated brackets. If
$D$ contains a dense set of separately analytic vectors for
$\big\{\pi(\xi_1),\ldots,\pi(\xi_k)\big\}$, then there exists a unique
unitary representation ${\Pi}$ of the connected simply connected
Lie group with Lie algebra ${\fg}$ on $\h$ such that
$d{\Pi}(\xi)\restriction D = {\pi(\xi)}$ for all $\xi \in {\fg}$.
\label{prop:fs}
\end{prop}

We recall that a vector $\psi$ is {\it
analytic\/} for an operator $X$ on $\h$ provided the series
\[\sum_{k=0}^{\infty}\frac{\|X^k\psi\|}{k!}\,t^k\]

\noi is defined and converges for some $t>0.$ If
$\{X_1,\ldots,X_k\}$ is a  set of operators defined on a common
invariant dense domain $D$, a vector $\psi 
\in D$ is {\it separately analytic\/} for $\{X_1,\ldots,X_k\}$ if
$\psi$ is analytic for each $X_j$. By a slight abuse of terminology,
we will say that a vector is separately analytic for a \la\ of
operators ${\cal X}$ if it is separately analytic for some  Lie
generating set
$\{X_1,\ldots,X_k\}$ of ${\cal X}$.

If it happens that $\fb$ is \id , then there need not exist a Lie
group having $\fb$ as its Lie algebra. Even if such a Lie group
existed, integrability is far from automatic, and technical
difficulties abound. Thus we will not insist that a quantization be
integrable in general. On the other hand, the analyticity requirement
in Proposition~\ref{prop:fs} makes sense under all
circumstances,\footnote{\,As long as $\fb$ is finitely generated,
which is assured by (B1).} and does guarantee integrability when
$\fb$ is \fd , so we will adopt it in lieu of integrability.

Finally, we will require that a quantization $\q$ be faithful on
$\fb$. While faithfulness is not usually assumed in the definition of
a quantization, it seems to us a reasonable requirement in that a
classical observable can hardly be regarded as ``basic'' in a
physical sense if it is in the kernel of a quantization map. In this
case, it cannot be obtained in any classical limit from a quantum
theory.

Therefore we have at last: 
\begin{defn} \, {\rm  A {\it quantization\/} of the pair
$({\oo},{\fb})$ is a prequantization $\q$ of $\oo$ on Op($D$)
satisfying
\begin{description}
\item \rule{0mm}{0mm}
\begin{enumerate}
\vspace{-4ex}
\item[(Q4)] $\q\restriction\fb$ is irreducible,  
\vskip 6pt
\item[(Q5)] $D$ contains a dense  set of separately analytic vectors
for $\q(\fb),$ and
\vskip 6pt
\item[(Q6)] $\q\restriction\fb$ is faithful.
\label{def:q}
\end{enumerate}
\end{description}}
\label{defn:quant}
\end{defn}

\noi \emph{Remarks.} 5. There are a number of analyticity assumptions
similar to (Q5) that one could make \cite{fl}; we have chosen the
weakest possible one.

6. (Q5) is not a severe restriction: When $\fb$ is
finite-dimensional, it is always possible to find representations of
it on domains $D$ which satisfy this property \cite{fl}. On the other
hand, nonintegrable representations do exist in general
\linebreak \cite[p.~247]{fl}.

7. Proposition~\ref{prop:fs} requires that a specific generating
set for $\q\big(\fb\big)$ be singled out. This also is not a
severe restriction: In examples, $\fb$ is often specified in this
manner. It is possible that (Q5) could be satisfied for one such set
but not another, but Remark 6 shows that the domain
$D$ can be chosen in such a way that this cannot happen if $\fb$
is finite-dimensional.

8. It is important to realize that irreducibility does not imply
integrability. For instance, there is an irreducible representation of
h(2) which is not integrable \cite[p.~275]{ReSi}.

\ms

We end this section with a brief comment on the domains $D$ appearing
in Definition~\ref{def:q}. For a representation $\pi$ of a Lie algebra
${\fg}$ on a Hilbert space $\h$, there is typically a multitude of
common, invariant dense domains that one can use as carriers of the
representation. (See \cite[\S 11.2]{b-r} for a discussion of some of
the possibilities.) But what is ultimately important for our purposes
are the closures $\ol{\pi(\xi)}$ for $\xi
\in {\fg}$, and not the
$\pi(\xi)$ themselves. So we do not want to distinguish between two
representations $\pi$ on Op$(D)$ and $\pi'$ on Op$(D')$ whenever
$\ol{\pi(\xi)} = \ol{\pi'(\xi)}$, in which case we say that $\pi$ and
$\pi'$ are {\it coextensive}. In particular, it may happen that the
given domain $D$ for a representation $\pi$ does not satisfy (Q5), but
there is an extension to a coextensive representation $\pi'$ on a
domain $D'$ that does.\footnote{\,A simple illustration is provided
by the Schr\"odinger representation (\ref{eq:srep}) with $D =
C^{\infty}_c(\r^n,\!\bc)$ and $D' =\s (\r^n,\!\bc)$.} In such cases we
will suppose that the representation has been so extended.

\end{section}


\begin{section}{Examples}

In this section we present the gist of the arguments---more or less
as they originally appeared---that there are no nontrivial
polynomial quantizations of either
$\r^{2n}$,
$S^2$, or $\T \!S^1$, with the \ba s h$(2n)$, su$(2)$, and e(2),
respectively. The complete proofs can be found in
\cite{a-m,Ch1,Fo,GGra,go80,go4,Gr,GS,vH1,vH2} for
$\r^{2n}$,
\cite{GGH} for $S^2$, and \cite{GG1} for $\T \!S^1$. The proofs in
all three examples require a detailed knowledge of the structure of
the Poisson algebras involved and their representations.
Finally, we show following
\cite{GGra} that there is a polynomial quantization of $\T \r_+$
with the basic algebra a(1), and following \cite{Go} that there is a
full quantization of
$T^2$ with the \ba\
$\ft$.
We also take this opportunity to repair a defect in the
standard presentations of the
\gr-\vh\ theorem for
$\r^{2n}.$

As an aside, we point out that many of the calculations in \S\S 5.2
and 5.3 were done using the {\sl Mathematica\/} package
\emph{NCAlgebra} \cite{HM}.


\begin{subsection}{$\r^{2n}$}

Before proceeding with the no-go theorem for
$\r^{2n}$, we remark that already at a purely mathematical level one
can observe a suggestive structural mismatch between the classical and
the quantum formalisms. Since a prequantization is essentially a Lie
algebra homomorphism, it ``compares'' the Lie algebra structure
of
$C^{\infty}(\r^{2n})$ with the Lie algebra of  (skew-) symmetric
operators (preserving a dense domain
$D$) equipped with the commutator bracket. But if we take $P
\subset C^{\infty}(\r^{2n})$ to be the subalgebra of polynomials,
Joseph
\cite{Jo} has shown that $P$ has outer derivations, but the enveloping
algebra of the Heisenberg algebra h(2$n$)---and hence that of the
Schr\"odinger representation thereof on $L^2(\r^n)$---has none. In
the next section we pursue this line of reasoning, and present another
such ``algebraic'' no-go theorem to the effect that a unital Poisson
algebra can never be realized as an associative algebra with the
commutator bracket.


In particular, one can see at the outset that it is impossible for a
prequantization to satisfy the ``product $\ra$ anti-commutator'' rule.
Taking $n=1$ for simplicity, suppose $\q$ were a prequantization of
the polynomial algebra $P$ for which
\begin{equation}  \q(fg) = {\textstyle \frac{1}{2}}\big(\q(f)\q(g) +
\q(g)\q(f)\big)
\label{eq:ac}
\end{equation}

\noi for all $f,g \in P.$ Take $f(p,q) = p$ and $g(p,q) = q$. Then 
\begin{eqnarray*} {\textstyle \frac{1}{4}}\big(\q(p)\q(q) +
\q(q)\q(p)\big)^2 \hspace{-1ex} & = & \hspace{-1ex} 
\q(pq)^2 \\ \rule{0ex}{3ex} & = & \hspace {-1ex} \q(p^2q^2)\; = \;
{\textstyle
\frac{1}{2}}\big(\q(p)^2\q(q)^2 +
\q(q)^2\q(p)^2\big).
\end{eqnarray*}

\noi Now by (Q1) we have $[\q(p),\q(q)] = -i\hbar I$, so  that the
L.H.S. reduces to
\[\q(q)^2\q(p)^2 - 2i\hbar\q(q)\q(p) - {\textstyle \frac{1}{4}}\hbar^2
I\]

\noi while the R.H.S. becomes
\[\q(q)^2\q(p)^2 - 2i\hbar\q(q)\q(p) - \hbar^2 I.\]

\noi As the product $\ra$ anti-commutator rule is equivalent to the
squaring \vn\ rule $\q(f^2) =\q(f)^2$, this contradiction also
shows that the latter is inconsistent with prequantization. Note that
the contradiction is obtained on quartic polynomials; there is no
problem if consideration is limited to observables which are at most
cubic.

This argument only used axiom (Q1) in the specific instance
$[\q(p),\q(q)] = -i\hbar I$. Consequently, one still obtains a
contradiction if one drops (Q1) and instead insists that $\q$ be
consistent with Schr\"odinger quantization (in which context this one
commutation relation remains valid, cf.~(\ref{eq:srep})). This {\em
manifest\/} impossibility of satisfying the product $\ra$
anti-commutator rule while being consistent with Schr\"odinger
quantization is one reason we have decided to focus on the Lie
structure as opposed to the associative structure of
$C^{\infty}(M)$. See \cite{a-b} for further results in this direction.

\ms

We now turn to the no-go theorem for $\r^{2n}$. We shall state the
main results for $\r^{2n}$ but, for convenience, usually prove them
only for
$n=1$. The proofs for higher dimensions are immediate generalizations
of these. In what follows $P  = \r [q^1,\ldots,q^n, 
p_1,\ldots,p_n]$; note that $P^1 \cong {\rm h}(2n)$, $P_2
\cong {\rm sp}(2n,\r)$, and $P^2 \cong {\rm hsp}(2n,\r)$.

We first observe that there {\em does\/} exist a quantization
$d\varpi$ of the pair
$(P^2,P^1)$. For
$n=1$ it is given by the familiar formul{\ae}
\begin{eqnarray} d\varpi(q)  = q, & d\varpi(1) = I, & d\varpi(p) =
-i\hbar
\frac{d}{dq} 
\label{eq:p1} \\ d\varpi(q^2) = q^2, & d\varpi(pq) = -i\hbar
{\displaystyle
\left (q \frac{d}{dq} + \frac{1}{2}\right )}, &
d\varpi(p^2) = -\hbar^2 \frac{d^2}{dq^{2}}
\label{eq:p2}
\end{eqnarray}

\noi on the domain $\s (\r,\!\bc) \subset L^2(\r).$ Properties
(Q1)--(Q3) and (Q6) are readily verified. (Q4) follows automatically
since the restriction of
$d\varpi$ to
$P^1$ is just the Schr\"od\-inger representation. For (Q5), we recall
the well-known fact that the Hermite functions form a dense set of
separately analytic vectors for $d\varpi(P^1)$. Since these functions
are also separately analytic vectors for $d\varpi(P_2)$
\cite[Prop.~4.49]{Fo}, the operator algebra
$d\varpi(P^2)$ is integrable to a unique representation $\varpi$ of
the universal cover ${\widetilde{\rm{HSp}}}(2n,\r)$ of
HSp$(2n,\r)$ (thereby justifying our notation
``$d\varpi$'').\footnote{\,This representation actually drops to the
double cover of HSp$(2n,\r)$, but we do not need this fact here.}
$\varpi$ is known as the ``extended metaplectic representation''; 
detailed discussions of it may be found in \cite{Fo,GS}.

We call $d\varpi$ the ``extended metaplectic quantization.'' It has
the following crucial property.
\begin{prop} The extended metaplectic quantization is the {\em
unique\/}  quantization of 
$\big({\rm hsp}(2n,\r),{\rm h}(2n)\big)$ which exponentiates to a
unitary representation of \linebreak[2] $\widetilde{\rm HSp}(2n,\r)$.
\label{prop:unique}
\end{prop}

By ``unique,'' we mean up to unitary equivalence and coextension of
representations (as explained at the end of \S4).

\ms

\noi {\it Proof.} Suppose $\q$ were another such quantization of
$\big({\rm hsp}(2n,\r),{\rm h}(2n)\big)$ on some domain $D$ in
a Hilbert space
$\h$. Then
$\q\big({\rm hsp}(2n,\r)\big)$ can be integrated to a representation
$\tau$ of ${\widetilde{\rm HSp}}(2n,\r)$, and (Q4) implies that
$\tau$, when restricted to H$(2n)
\subset {\widetilde{\rm HSp}}(2n,\r)$, is irreducible. The Stone-\vn\
Theorem then states that this representation of H$(2n)$ is unitarily
equivalent to the Schr\"odinger representation, and hence $\tau =
U\varpi U^{-1}$ for some unitary map $U:L^2(\r^n) \ra \h$ by
\cite[Prop.~4.58]{Fo}. Consequently, $\q(f) = U\overline{d\varpi(f)}
U^{-1} \restriction D$ for all $f \in {\rm hsp}(2n,\r)$. Since the
Hamiltonian vector fields of such
$f$ are complete, the corresponding operators
$\q(f)$ are e.s.a., and therefore
$\overline{\q(f)} = U\overline{d\varpi(f)} U^{-1}$. Thus $\q(f)$ and
$U{d\varpi(f)} U^{-1}$ are coextensive.
\endproof

\ms 

The weakest version of the no-go theorem is:
\begin{thm}[Weak No-Go Theorem] The extended metaplectic quantization
of\/
$(P^2, P^1)$ cannot be extended beyond $P^2$ in $P$.
\label{thm:weak}
\end{thm}

Since $P^2$ is a maximal Lie subalgebra of $P$ \cite[\S 16]{GS},
we may restate this as: {\em There exists no quantization of\/
$(P,P^1)$ which reduces to the extended metaplectic quantization on
$P^2.$}
\ms

\noi {\it Proof.} Let $\q$ be a quantization of $(P,P^1)$ which
extends the metaplectic quantization of $(P^2,P^1)$. We will show that
a contradiction arises when cubic polynomials are considered.

By inspection of (\ref{eq:p1}) and (\ref{eq:p2}), we see that the
product
$\ra$ anti-commutator rule (\ref{eq:ac}) is valid for $f,\,g \in P^1.$
In particular, we have the \vn\ rules 
\begin{equation}
\q(q^2)=\q(q)^2,\;\; \q(p^2)=\q(p)^2 
\label{eq:vnrr2n1}
\end{equation}

\noi and
\begin{equation}
\q(qp) = {\textstyle \frac{1}{2}}\big(\q(q) \q(p) + \q(p) \q(q)\big). 
\label{eq:vnrr2n2}
\end{equation} 

\noi These in turn lead to higher degree \vn\ rules.

\begin{lem} For all real-valued polynomials $r$,
\[\q\big(r(q)\big) = r\big(\q(q)\big),\;\;\;\q\big(r(p)\big)=
r\big(\q(p)\big),\]
\[\q\big(r(q)p\big) = {\textstyle
\frac{1}{2}}\big[r\big(\q(q)\big)\q(p)+\q(p)r\big(\q(q)\big)\big],\]

\noi and 
\[\q\big(qr(p)\big) = {\textstyle
\frac{1}{2}}\big[\q(q)r\big(\q(p)\big)+r\big(\q(p)\big)\q(q)\big].\]
\label{lem:morevnrs}
\end{lem}

\vskip -2.5ex

\noi {\it Proof.} We illustrate this for $r(q) = q^3$. The other rules
follow similarly using induction. Now $\{q^3,q\} = 0$ whence by (Q1)
we have
$[\q(q^3),\q(q)] = 0$. Since also
$[\q(q)^3,\q(q)] = 0$, we may write $\q(q^3) =
\q(q)^3 + T$ for some operator $T$ which (weakly) commutes with
$\q(q)$. We likewise have using (\ref{eq:vnrr2n1})
\[ [\q(q^3),\q(p)] =  -i\hbar\,
\q(\{q^3,p\}) 
 =  3i\hbar\,\q(q^2) =  3i\hbar\,\q(q)^2 =
[\q(q)^3,\q(p)]\]

\noi from which we see that $T$ commutes with $\q(p)$ as well.
Consequently, $T$ also commutes with $\q(q)\q(p)+ \q(p)\q(q)$. But
then from (\ref{eq:vnrr2n2}),
\begin{eqnarray*}\q(q^3) & = & {\textstyle \frac{1}{3}}\,
\q\big(\{pq,q^3\}\big) = {\textstyle
\frac{i}{3\hbar}}\,[\q(pq),\q(q^3)] \\ \rule{0ex}{3ex} & =&
{\textstyle
\frac{i}{3\hbar}}\,\left[{\textstyle
\frac{1}{2}}\big(\q(q)\q(p) +
\q(p)\q(q)\big),\q(q)^3 + T\right] \\ \rule{0ex}{3ex} & = &
{\textstyle
\frac{i}{6\hbar}}\,[\q(q)\q(p) +
\q(p)\q(q),\q(q)^3] 
 = \q(q)^3.  
\end{eqnarray*}
 \vskip -5ex \hfill $\bigtriangledown$

\ms
\vskip 1.5ex
\ms

With this lemma in hand, it is now a simple matter to prove the no-go
theorem. Consider the classical equality 
\[{\textstyle \frac{1}{9}}\{q^3,p^3\}= {\textstyle
\frac{1}{3}}\{q^2p,p^2q\}.\]

\noi Quantizing and then simplifying this, the formul{\ae} in
Lemma~\ref{lem:morevnrs} give
\[\q(q)^2\q(p)^2 - 2i\hbar\q(q)\q(p) - {\textstyle
\frac{2}{3}}\hbar^2 I\]

\noi for the L.H.S., and 
\[\q(q)^2\q(p)^2 - 2i\hbar\q(q)\q(p) - {\textstyle \frac{1}{3}}\hbar^2
I\]

\noi for the R.H.S., which is a contradiction.
\endproof

\ms


In \gr's paper \cite{Gr} a stronger result was claimed; in our
terminology, his assertion was that there is no quantization of
$(P,P^1)$. This is \emph{not} what Theorem~\ref{thm:weak} states. For
if
$\q$ is a quantization of $(P,P^1)$, then while of course $\q(P^1)$
must coincide with Schr\"odinger quantization, it is not obvious
that $\q$ need be the extended metaplectic quantization
when restricted to
$P^2$. The problem is that $\q(P^2)$ is not \emph{a priori}
integrable; (Q5) only guarantees that
$\q(P^1)$ can be integrated.

\vh\ supplied an extra assumption guaranteeing the
integrability of $\q(P^2)$, which in particular implies: If the
Hamiltonian vector fields of $f,\,g$ are complete and $\{f,g\} = 0$,
then $\q(f)$ and $\q(g)$ {\em strongly\/} commute \cite{vH1}. This
assumption is used to derive the \vn\ rules (\ref{eq:vnrr2n1}) and
(\ref{eq:vnrr2n2}) in \cite{a-m, Ch1}. It is also possible to enforce
integrability in a more direct manner \cite{GGT}. 

It turns out that it is possible to establish the integrability of
$\q(P^2)$ \emph{without} introducing extra assumptions, via the
following generalization of Proposition~\ref{prop:unique}
\cite{go4}.
\begin{prop}
\label{prop:fix}
Up to coextension of representations, any quantization of
\linebreak
$(P^2,P^1)$ is unitarily equivalent to the extended metaplectic
quantization.
\end{prop}

\noi {\it Proof.} Let $\q$ be a quantization of $(P^2,P^1)$. Arguing
as in the proof of Proposition~\ref{prop:unique}, we may assume that
$\q(P^1)$ is the Schr\" odinger representation (\ref{eq:srep}). We
will prove by brute force that the \vn\ rules (\ref{eq:vnrr2n1}) and
(\ref{eq:vnrr2n2}) hold. Again taking $n=1$, we give the details for
$q^2$; the calculations for $qp$ and $p^2$ are similar.

Writing $|k\rangle = h_k(q)$, from (\ref{eq:hermite}) we compute 
\begin{equation}
\q(q) |k\rangle = k|k-1\rangle + \textstyle{\frac{1}{2}}|k+1\rangle
\;\; \mbox{and} \;\;
\q(p) |k\rangle = -i\hbar \big(k|k-1\rangle -
\textstyle{\frac{1}{2}}|k+1\rangle \big).
\label{eq:kets}
\end{equation}

\noi Since the Hermite functions $\{\,|k\rangle\,|\,k = 0,1,\ldots\}$
form a basis of
$L^2(\r)$, we may expand $\q(q^2)|k\rangle =
\sum_{j=0}^\infty E_{k,j}|j\rangle$. Now using (\ref{eq:kets}),
we compute the matrix elements
$\langle j|\,[\q(q),\q(q^2)]\,|k\rangle = 0$ to get
\begin{equation*}
(j+1)E_{k,j+1} + \textstyle{\frac{1}{2}}E_{k,j-1} - kE_{k-1,j} -
\textstyle{\frac{1}{2}}E_{k+1,j} = 0.
\end{equation*}

\noi Similarly, the identity $\langle
j|\,[\q(p),\q(q^2)]\,|k\rangle = -2i\hbar\langle
j|\q(q)|k\rangle$ reduces to
\begin{equation*}
(j+1)E_{k,j+1} - \textstyle{\frac{1}{2}}E_{k,j-1} - kE_{k-1,j} +
\textstyle{\frac{1}{2}}E_{k+1,j} = 2(j+1)\delta_{j,k-1} +
\delta_{j,k+1}.
\end{equation*}

\noi Subtracting and adding these two equations produce
\begin{eqnarray}
E_{k+1,j} - E_{k,j-1} & = & 2(j+1)\delta_{j,k-1} +
\delta_{j,k+1}
\label{eq:me1} \\ \rule{0ex}{3ex}
2(j+1)E_{k,j+1} - 2kE_{k-1,j} & = & 2(j+1)\delta_{j,k-1} +
\delta_{j,k+1}. \nonumber
\end{eqnarray}

\noi In the last equation, reindex $j \mapsto j-1$ and $k \mapsto
k+1$:
\begin{equation}
\label{eq:me2} 
2jE_{k+1,j} - 2(k+1)E_{k,j-1}  =  2j\delta_{j-1,k} +
\delta_{j-1,k+2}.
\end{equation}

\noi Solve (\ref{eq:me1}) and (\ref{eq:me2}) simultaneously to get 
\begin{equation*}
E_{k,j-1} = \frac{2j(j+1)\delta_{j,k-1}-
\frac{1}{2}\delta_{j,k+3}}{k+1-j}.
\end{equation*}

\noi Again reindexing $j \mapsto j+1$, we have finally
\begin{equation}
E_{k,j} = \frac{2(j+1)(j+2)\delta_{j,k-2}-
\frac{1}{2}\delta_{j,k+2}}{k-j}.
\label{eq:me3}
\end{equation}

\noi This yields
\begin{equation}
E_{k,k+2} = 1/4 \;\;\; \mbox{and} \;\;\; E_{k,k-2} = k(k-1);
\label{eq:me4}
\end{equation}

\noi all other $E_{k,j}= 0$ with the exception of $E_{k,k}$, which is
not determined by (\ref{eq:me3}). However, taking $j = k+1$ in
(\ref{eq:me1}) gives $E_{k+1,k+1} - E_{k,k} = 1$, whence
$E_{k,k}=E_{0,0} + k.$

Comparing this and (\ref{eq:me4}) with
\[\q(q)^2|k\rangle = k(k-1)|k-2\rangle + \left(k +
\textstyle{\frac{1}{2}}\right)|k\rangle + \textstyle{\frac{1}{4}}
|k+2\rangle,\]

\noi we conclude that $\q(q^2) = \q(q)^2 + EI$ for some (real)
constant $E.$ Likewise, $\q(p^2) = \q(p)^2 + FI$ and $\q(qp) =
\frac{1}{2}\big(\q(q) \q(p) + \q(p) \q(q)\big) + GI.$ But then
\begin{eqnarray*}
2i\hbar \big(\q(q) \q(p) + \q(p)\q(q)\big) \hspace{-1ex} &= &
\hspace{-1ex} [\q(q)^2,\q(p)^2] \; = \; 
 [\q(q^2),\q(p^2)] \\ \rule{0ex}{3ex}
& =  & \hspace{-1ex} -i\hbar \q(\{q^2,p^2\}) \; = \; 4i\hbar \q(qp),
\end{eqnarray*}

\noi which implies that $G=0$. Similarly $E=F=0.$
\endproof
 
\ms

If $\q$ were a quantization of $(P,P^1)$, $\q(P^2)$ must therefore be
unitarily equivalent to the extended metaplectic quantization, and
this contradicts Theorem~\ref{thm:weak}. Thus we have proven our main
result:
\begin{thm}[Strong No-Go Theorem] There is no quantization
of $(P,P^1)$.
\label{thm:strong}
\end{thm}

Van~Hove \cite{vH1} gave a slightly different analysis using only
those observables $f\in C^\infty(\r^{2n})$ with complete Hamiltonian
vector fields, and still obtained an obstruction (but now to
quantizing all of $C^\infty(\r^{2n})$). Yet another variant of \gr 's
Theorem will be presented in \S 6.5.

\ms

We hasten to add that there are  subalgebras of $P$ other
than $P^2$ which can be quantized. For example, let 
\[S = \left\{\sum_{i=1}^n f^i(q)p_i + g(q)\right\},\]

\noi where $f^i$ and $g$ are polynomials. Then it is straightforward
to verify that for each $\eta \in \r,$ the map $\q_{\eta}:S \ra
{\rm Op}\big(\s (\r^n,\!\bc)\big)$ given by 
\begin{equation}
\q_{\eta}\left(\sum_{i=1}^n f^i(q)p_i + g(q)\right) =
-i\hbar \sum_{i=1}^n\left(f^i(q)\frac{\partial}{\partial q^i} +
\left[
\frac{1}{2} + i\eta \right]\frac{\partial f^i}{\partial q^i}\right) +
g(q)
\label{eq:sigma}
\end{equation}

\noi is a quantization of $(S,P^1).$ $\q_0$
is the familiar ``position'' or ``coordinate representation.'' The
significance of the parameter $\eta$ is explained in \cite{ADT}. Since
$S$ is also a maximal subalgebra of $P$, (Q1) implies that any
quantization which extends $\q_\eta$ must be defined on all of
$P$. Thus Theorem~\ref{thm:strong} yields
\begin{cor} The quantizations $\q_\eta$ of \
$(S,P^1)$ cannot be extended beyond
$S$ in $P$.
\end{cor}

The cases when $\eta \neq 0$ will be studied further in \S 5.3. We
limit ourselves to pointing out that Proposition 7 in \cite{go4}
yields ``uniqueness'': Any quantization of $(S,P^1)$ must
be of the form $\q_{\eta}$ for some $\eta \in \r.$

 A similar analysis applies to the ``Fourier transform'' of the
subalgebra
$S,$ i.e., the ``momentum'' subalgebra of all polynomials which are at
most affine in the coordinates $q^i.$ In fact, it turns out 
(at least for $n = 1$) that $P^2$ and
$S$ exhaust the list of isomorphism classes of maximal
Lie subalgebras of $P$ which contain $P^1$ \cite{go4}.

\end{subsection}


\begin{subsection}{$S^2$} 

Now we turn our attention to the sphere.
Since $S^2$ is compact, all classical observables are complete.
Moreover, su(2) $\cong$ span$\{S_1,S_2,S_3\}$ is a
compact simple Lie algebra. Consequently the functional analytic
subtleties present in the case of
$\r^{2n}$ disappear. But the actual computations, which were fairly
routine for
$\r^{2n}$, turn out to be much more complicated for
$S^2.$ 
 
The Poisson bracket on $C^{\infty}(S^2)$ corresponding to
(\ref{eq:sfs2}) is 
\[\{f,g\}= - \sum_{i,j,k = 1}^3\epsilon_{ijk}\,S_i\,\frac{\partial
f}{\partial S_j} \frac{\partial g}{\partial S_k}.\]

\noi In particular, we have the relations
$\{S_j,S_k\} = -\sum_{l = 1}^3\epsilon_{jkl}\,S_l$. In this example
$P$ is the polynomial algebra $\r[S_1,S_2,S_3]$ in the components
of the spin vector, subject to the relation
\begin{equation}
S_1\,\!^2 + S_2\,\!^2 + S_3\,\!^2 = s^2.
\label{eq:s2casimir}
\end{equation}

\noi We may identify
$P$ with the space of spherical harmonics. We have $P_1 \cong
{\rm su}(2)$ and $P^1 \cong {\rm u}(2)$. 

\ms

Let $\q$ be a quantization of $(P,P_1)$ on a
Hilbert space $\h$, whence
\begin{equation}
[\q(S_j),\q(S_k)] = i\hbar \sum_{l=1}^3
\epsilon_{jkl}\,\q(S_l)
\label{eq:cr}
\end{equation}

\noi and
\begin{equation}
\q({\bf S}^2) = s^2I.
\label{eq:cc}
\end{equation}

\noi By (Q5) and Proposition~\ref{prop:fs}, $\q({\rm su}(2))$
can be exponentiated to a unitary representation of SU(2) which,
according to (Q4), is irreducible. Therefore $\h$ must be
finite-dimensional, and
$\q({\rm su}(2))$  must be one of the usual spin angular
momentum representations, labeled by $j=0,\,\hlf,\,1,\ldots$ For a
fixed value of
$j$, $\dim \h = 2j+1$ and
\begin{equation}
\sum_{i=1}^3\q(S_i)^2=\hbar^2j(j+1)I.
\label{eq:qc}
\end{equation}

Our goal is show that no such (nontrivial) quantization exists.
Patterning our analysis after that for $\r^{2n}$, we use
irreducibility to derive some 
\vn\ rules.
\begin{lem} For $i = 1,\,2,\,3$ we have
\begin{equation} 
\q(S_i\,\!^2)=a\q(S_i)^2+cI
\label{eq:sisi}
\end{equation}

\noi where $a,\,c$ are representation dependent real constants with
$a^2 + c^2 \neq 0$.
\label{lem:vnrs21}
\end{lem}

The proof is in \cite{GGH}. {}From this we also derive
\begin{equation}
\q(S_iS_k)={\displaystyle
\frac{a}{2}}\big(\q(S_i)\q(S_k)+\q(S_k)\q(S_i)\big)
\label{eq:sisk}
\end{equation}

\noi for $i \neq k$. (As an aside, these formul{\ae} show that a
quantization, if it exists, may be badly behaved with respect to the
multiplicative structure on
$C^{\infty}(S^2)$; in particular, the product $\ra$ anti-commutator
rule need not hold. Remarkably, this is as it should be: For {\em
if\/} this rule were valid, then -- subject to a few mild assumptions
on $\q$---the classical spectrum of $S_3$, say, would have to
coincide with that of $\q(S_3)$ which is contrary to experiment
\cite{GGH}.) With these tools, we can now prove the main result: 
\begin{thm} There is no nontrivial quantization of $(P,P_1)$.
\label{thm:nogos2}
\end{thm}

\noi {\it Proof.}  
Fix $j>0$, as $j=0$ produces a trivial quantization. Assuming that
$\q$ is a quantization of $(P,P_1)$, we can use
(\ref{eq:cr})-(\ref{eq:sisk}) to quantize the classical relation
\begin{equation}
s^2S_3=\big\{S_1\,\!^2-S_2\,\!^2,\,S_1S_2\big\}-
\big\{S_2S_3,\,S_3S_1\big\},
\label{eq:s2pbr}
\end{equation}

\noi thereby obtaining
\begin{equation} s^2=a^2\hbar^2\big(j(j+1)-{\textstyle
\frac{3}{4}}\big)
\label{eq:xxx}
\end{equation}

\noi which  contradicts $s>0$ for $j=\hlf$. Now assume $j>\hlf$, and
quantize
\begin{equation} 
2s^2S_2S_3=\big\{S_2\,\!^2,\{S_1S_2,S_1S_3\}\big\} -
{\textstyle \frac{3}{4}}
\big\{S_1\,\!^2,\{S_1\,\!^2,S_2S_3\}\big\},
\label{eq:s2pbr2}
\end{equation}

\noi similarly obtaining
\[s^2=a^2\hbar^2\big(j(j+1)-{\textstyle
\frac{9}{4}}\big)\]

\noi which contradicts (\ref{eq:xxx}). Thus we have derived
contradictions for all $j>0$, and the theorem is proven.
\endproof

\ms

In view of the impossibility of quantizing $(P,P_1)$, one can ask what
the maximal Lie subalgebras in $P$ are to which we can extend an
irreducible representation of $P_1$. The following chain of results,
which we quote without proof (cf. \cite{GGH}), provides the answer.
\begin{prop}
$P^1$ is a maximal Lie subalgebra of
$\r \oplus O\subset P$, where $O$ is the 
\linebreak
Poisson algebra consisting of
polynomials containing only terms of odd degree.
\label{prop:maxx}
\end{prop}

Next we establish a no-go theorem for $(\r \oplus O,P_1)$. However,
the Von~Neumann rules listed in Lemma~\ref{lem:vnrs21}
involve only even degree polynomials, so these are not applicable in
$O$. Fortunately, we have another set of Von~Neumann
rules, also implied by the irreducibility of
$\q(P_1)$, involving only terms of odd degree.
\begin{lem} If $\q$ is a quantization of $(\r \oplus O,P_1)$, then for
$i=1,\,2,\,3$,
\[\q(S_i\,\!^3)=b\q(S_i)^3+e\q(S_i)\]

\noi where $b,\,e\in\r$.
\end{lem}

{}From this we prove (with far greater effort): 
\begin{thm} There is no nontrivial quantization of
$(\r \oplus O,P_1)$.
\label{thm:yyy}
\end{thm}

Now $\r \oplus O$ is itself a maximal Lie subalgebra of $P$, and in
fact the only Lie subalgebras of $P$ strictly containing $P_1$
are $P^1$,
$\r \oplus O$, and $P$ itself. On the other hand, $P^1 = \r
\oplus P_1$ is obviously quantizable. Thus Theorem~\ref{thm:yyy} and
Proposition~\ref{prop:maxx} combine to yield our sharpest result for
the sphere:
\begin{cor} No nontrivial quantization of $(P^1,P_1)$ can be extended
beyond $P^1$ in $P$.
\end{cor}

Thus within the algebra of polynomials, \big(u(2),su(2)\big) is
the most one can quantize.

There are crucial structural differences between the
Groenewold-Van~Hove analysis of $\r^2$ and the current analysis of
$S^2$. Within $P = \r[q,p]$ the Heisenberg algebra has as its
Lie normalizer the algebra of polynomials of degree at most 2,
and there is no obstruction to quantization in this algebra: The
obstruction comes from the cubic polynomials. On the other hand, for
the sphere, the special unitary algebra 
has as its normalizer the algebra of polynomials of degree at most
one; we obtain an obstruction in the quadratic polynomials, and find
that there is no extension possible for a quantization of $P^1$. The
fact that this su(2)-subalgebra is essentially self-normalizing is one
reason why we are able to obtain ``strong'' no-go results for the
sphere relatively easily (as compared to
$\r^{2n}$).

\end{subsection}


\begin{subsection}{$\T \! S^1$} 

Our final example of an obstruction is provided by the symplectic
cylinder, which appears in geometric optics \cite[\S 17]{GS}.
Endow
$\T \! S^1$ with the canonical Poisson bracket
\[\{f,g\} = \frac{\partial f}{\partial \ell}\frac{\partial
g}{\partial \theta} - \frac{\partial f}{\partial \theta}\frac{\partial
g}{\partial \ell},\] 

\noi where $\ell$ is the angular momentum conjugate to $\theta$.
While the symplectic self-action of
$T^*\!S^1$ is not Hamiltonian (thinking of
$T^*\!S^1$ as $S^1 \! \times
\r$), the cylinder can nonetheless be
realized as a coadjoint orbit of the special Euclidean group
SE(2) \cite[\S 14.8]{m-r}.
The corresponding momentum map
$T^*\!S^1 \ra \mbox{e}(2)^*$ has components $\{\sin \theta,
\cos \theta,\ell\}$; therefore we take as a basic algebra
\[{\rm e}(2) \cong \sp\{\sin \theta, \cos \theta, \ell\}.\] 

\noi The polynomial algebra $P$ generated by this \ba\ consists of
sums of multiples of terms of the form $\ell^{\, r} \sin^m\!\theta
\cos^n\!\theta$ of total degree $r+m+n$ with
$r,\,m,\,n$ nonnegative integers.  Then $P_1 \cong {\rm e}(2)$. 

Our first task is to determine all possible quantizations of the
basic algebra e(2). By virtue of (Q3) and (Q4), for this it
suffices to compute the derived representations corresponding to the
IURs of the universal
covering group of
${\mbox{SE(2)}}$, which is the semidirect
product $\r \ltimes \r^2$ with the composition law 
\[(t,x,y)\cdot
(t',x',y') = (t+t', x' \cos t + y' \sin t + x, y' \cos t - x' \sin
t + y).\]

\noi {}From the theory of induced representations of
semidirect products \cite{Ma} (see also \cite[\S5.8]{Is}), we
find that the only nontrivial IURs are \id ; up to unitary
equivalence they take the form
\[\big(U(t,x,y)\psi\big)(\theta) = e^{i\lambda(x \cos
\theta + y
\sin \theta)}e^{i\nu t}\psi(\theta + t) \]

\noi on
$L^2(S^1).$ Here $\lambda, \nu$ are real parameters satisfying
$\lambda > 0$ and $0 \leq \nu < 1.$ We identify $\lambda$ with the
reciprocal of $\hbar$, cf.
\cite{GG2,Is}. After rescaling appropriately, the corresponding
derived representations become
\begin{equation} \q(\ell) = -i\hbar\left({\displaystyle
\frac{d}{d\theta}} + i\nu I \right), \;\;
\q(\sin
\theta) =  \sin \theta, \;\; \q(\cos \theta) =  \cos
\theta
\label{eq:(i)'}
\end{equation}

\noi on 
$C^{\infty}(S^1,\!\bc)$. 

Just as with our previous examples, we use irreducibility to obtain
\vn\ rules. In \cite{GG2} we compute
\[\q(\ell^{\, 2}) = \q(\ell)^2 + cI,\]

\noi where $c \in
\r$ is arbitrary. {}From this and (\ref{eq:(i)'}) we eventually derive
\begin{equation}
\q(\ell^{\, 2} \sin \theta) =
\q(\sin \theta)\q(\ell)^2 -i\hbar\q(\cos \theta)\q(\ell) +
\frac{\hbar^2}{4}\q(\sin \theta)
\label{eq:l2s}
\end{equation}

\vskip -1ex

\noi and

\vskip -3ex

\begin{equation}
\q(\ell^{\, 2} \cos \theta) =
\q(\cos \theta)\q(\ell)^2 + i\hbar\q(\sin \theta)\q(\ell) +
\frac{\hbar^2}{4}\q(\cos
\theta).
\label{eq:l2c}
\end{equation}

\begin{thm} There is no nontrivial quantization of $(P,P_1)$.
\label{thm:dnecyl}
\end{thm}

\noi {\em Proof.} We merely use (\ref{eq:(i)'})--(\ref{eq:l2c}) to
quantize the bracket relation
\begin{equation}
2\big\{\{\ell^{\, 2} \sin \theta,\ell^{\, 2} \cos
\theta\},
\cos
\theta\big\} = 12\ell^{\, 2} \sin \theta.
\label{eq:T*s1pbr}
\end{equation}

\noi After simplifying, the left hand side reduces to
\[ 12\q(\sin \theta)\q(\ell)^2 - 12i\hbar\q(\cos \theta)\q(\ell) + 
5\hbar^2\q(\sin \theta),\]

\noi whereas the right hand side is
\[ 12\q(\sin \theta)\q(\ell)^2 - 12i\hbar\q(\cos \theta)\q(\ell) + 
3\hbar^2\q(\sin \theta),\]

\noi and the required contradiction is evident. \endproof

\ms

We next determine the maximal Lie subalgebras of $P$ to which we can
extend an irreducible representation of $P_1$. Such subalgebras
certainly exist: For instance, there is a two-parameter family of
quantizations of the pair
$(L^1,P_1)$, where $L^1$ denotes the Lie subalgebra of
polynomials which are at most first degree in $\ell$. They are the
``position representations'' on
$C^{\infty}(S^1,\!\bc) \subset L^2(S^1)$ given by 
\begin{equation}
\q_{\nu,\eta}\big(f(\theta)\ell + g(\theta)\big) =
-i\hbar\left(f(\theta)\frac{d}{d\theta} +
\left[\frac{1}{2} + i\eta\right]f'(\theta) + i\nu
f(\theta)\right) + g(\theta),
\label{eq:qnu}
\end{equation}

\noi where $\nu$ labels the IURs of the universal cover of SE(2) and
$\eta$ is real. 

To this end we classify the maximal Lie subalgebras of $P$ containing
$P_1$. For each $\alpha \in \r$ let $V_\alpha$ be the Lie subalgebra
generated by
\[\{1,\; \sin \theta,\; \cos \theta,\; \ell, \; \ell(\ell +
\alpha)\cos \,(2N+1)\theta,\;\ell(\ell +
\alpha)\sin \,(2N+1)\theta \,|\,N \in \bf N\}.\]

\noi Although far from obvious, it turns out that \cite{GG2}
\begin{prop}
$L^1$ and $V_\alpha, \alpha \in \r,$ are the only proper maximal Lie
subalgebras of
$P$ strictly containing $P_1$.
\label{prop:max}
\end{prop}

In contrast to
$L^1$, it is possible to show that there is {\em no\/}
nontrivial quantization of any $V_\alpha$ which represents $P_1$
irreducibly. (While the method of proof is the same as that of the
no-go theorem for $P$ presented above, we must make sure that all
constructions take place in $V_\alpha$. The details may be found in
\cite{GG2}.) Since $L^1$ is
maximal, Theorem~\ref{thm:dnecyl} implies
that none of the quantizations $\q_{\nu,\eta}$ can be extended beyond
$L^1$ in $P.$ Furthermore \cite{GG2},
the quantizations (\ref{eq:qnu}) of $L^1$ are the only possible ones:
\begin{thm} If $\q$ is a nontrivial quantization of $(L^1,P_1)$,
then $\q = \q_{\nu,\eta}$ for some
$\nu \in [0,1)$ and $\eta \in \r$.
\label{thm:unique}
\end{thm}

Taken together, these results completely characterize the polynomial
quantizations for the \ba\ e(2).

\ms

Since $\T\! S^1$ is covered by $\r^2$, and as e(2) is the
natural analogue for the cylinder of h(2) for the plane, the
quantization of the former might be expected to share some of the
features of that of the latter, and we see from the above that in most
respects this is so. In both examples there is an obstruction, and a
maximal Lie subalgebra of polynomial observables that can be
consistently quantized consists of those polynomials which are affine
in the momentum. 

There are some differences, however, which reflect the non-simple
connectivity of $T^*\!S^1$. For instance, on $\r^2$, there are
exactly two isomorphism classes of maximal polynomial Lie
subalgebras containing the basic algebra
$\sp\{1,q,p\}$, whereas according to Proposition~\ref{prop:max}
there are also two such containing 
\linebreak
$\{\sin \theta,\cos
\theta,\ell\}$ for the cylinder. However, on $\r^2$ all of
these maximal subalgebras can be consistently quantized, but on
$T^*\!S^1$ only one of these can (viz. $L^1$). (Since $P^2$ is not a
\lsa\ of $P$, there is no analogue of the
metaplectic representation for $\T \!S^1$ and, since
$\theta$ is an angular variable, there is also no cylindrical
counterpart of the momentum representation.) Thus the
possible polynomial quantizations of $T^*\!S^1$ are more limited than
those of $\r^2$.

One topic for future exploration would be to consider the
higher-dimensional analogues of the cylinder, viz. $T^*\!S^n$ with
\ba\ e($n$).

\end{subsection}


\begin{subsection}{$\T \r_+$} 

We have encountered obstructions to quantization in the three examples
presented so far, despite the fact that $\r^{2}$, $\T \!S^1$, and
$S^2$ are quite different structurally. Topologically these
phase spaces range from contractible to compact, and algebraically
the \ba s h(2), e(2), and su(2) are nilpotent, solvable, and simple,
respectively. Moreover, the representations of these
algebras were in some instances unique and in others not, and they
were finite- as well as \id. This wide array of behaviors strongly
suggests that such obstructions should be ubiquitous. Therefore it
comes as a surprise that this is \emph{not} so
\cite{GGra}: there is no obstruction to polynomially quantizing $\T
\r_+ = \{(q,p) \in
\r^2\,|\,q >0\}$ with the ``affine'' \ba\
\[{\rm a}(1) \cong {\rm span}\{pq,q^2\}.\] 

Upon writing
\[X = pq, \;\;Y = q^2\]

\noi the bracket relation becomes $\{X,Y\} = 2Y.$ Thus a(1) is
the simplest example of a solvable algebra which is not
nilpotent. The simply connected covering group of
a(1) is isomorphic to the group ${\rm A}_+(1) = \r \rtimes \r_+$ of
orientation-preserving affine transformations of the line (hence the
terminology). It is straightforward to check that $\T \r_+$ with the
canonical \pb\ can be realized as a coadjoint orbit in a(1)$^*$
\cite[\S 14.1(b)]{m-r}.

The corresponding polynomial algebra $P = \r[X,Y]$ is free, and has
the crucial feature that for each $k \geq 0$, the subspaces $P_k$ are
\emph{ad}-invariant, i.e.,
\begin{equation}
\{P_1,P_k\} \subset P_k.
\label{ad}
\end{equation}

\noi (Note that $P_1 \cong {\rm a}(1)$). Because of this $\{P_k,P_l\}
\subset P_{k+l},$ whence each $P_{(k)}$ is a Lie ideal. We thus have
the semidirect sum decomposition
\begin{equation}
P = P^1 \ltimes P_{(2)}.
\label{decomp}
\end{equation}

Now on to quantization. Since
$P_{(2)}$ is a Lie ideal, we can obtain a quantization $\q$ of
\emph{all} of $P$ simply by finding an appropriate representation of
$P^1 = \r \oplus
P_1$ and setting
$\q(P_{(2)}) = \{0\}$!

Since A$_+$(1) is a semidirect product we can
generate the required representation of $P_1$ by induction.
Following the recipe in \cite[\S 17.1]{b-r} we obtain the
one-parameter family of unitary representations $U_{\pm}$ of A$_+$(1)
on $L^2(\r_+,dq/q)$ given by
\[\big(U_{\pm}(\nu,\lambda)\psi\big)(q) = e^{\pm i\mu \nu
q^2}\psi(\lambda q)\]

\noi with $\mu > 0.$ (As in the previous subsection, we identify
the parameter
$\mu$ with $\hbar^{-1}$.) According to
Theorems 4 and 5 in
\cite[\S 17.1]{b-r} these two representations (one for each choice of
sign) are irreducible and inequivalent; moreover, these are the
\emph{only} irreducible 
\id\ unitary ones. 

Writing
$\varrho_{\pm} =-i\hbar\, dU_{\pm}$ we get the 
quantization(s) of a(1) on
$L^2(\r_+,dq/q)$:
\[\varrho_{\pm}(pq) = -i\hbar q\frac{d}{dq},\;\;\varrho_{\pm}(q^2)
= \pm q^2.\]

\noi Extend these to $P^1$ by demanding that $\varrho_{\pm}(1) =
I$, and set
$\q_{\pm} =
\varrho_{\pm}
\oplus 0$ (cf. (\ref{decomp})). This is clearly a
prequantization of
$P$, by construction (Q4) and (Q5) are satisfied, and
$\q_{\pm}\restriction {\rm a}(1) = \varrho_{\pm}$ is clearly
faithful.  Thus $\q_{\pm}$
are the required quantization(s) of $(P,P_1)$.

\ms
\noi \emph{Remarks.} 9. The $+$ quantization of a(1) is exactly what
one obtains by geometrically quantizing $M$ in the vertical
polarization. Carrying this out, one gets $H = L^2(\r_+,dq)$ and
\[pq \mapsto -i\hbar\left(q\frac{d}{dq} +
\frac{1}{2}\right),\;\;q^2 \mapsto q^2.\]

\noi  The $+$ quantization is equivalent to this via the
unitary transformation 
\linebreak
$L^2(\r_+,dq/q) 
\rightarrow
L^2(\r_+,dq)$ which takes $f(q) \mapsto f(q)/\sqrt{q}.$ We do
not know if the $-$ quantization can be gotten via geometric
quantization. 

10. Note that
${\rm a}(1) \subset \mbox{sp}(2,\r)$. In fact, the $+$ and $-$
quantizations are equivalent to
the restrictions to a(1) of the metaplectic
quantizations on
$L^2_{\mbox{\scriptsize even}}(\r,dq)$ and
$L^2_{\mbox{\scriptsize odd}}(\r,dq)$ (cf. \S 5.1 and Remark 9). 

11. Since $\q(P_{(2)}) = 0$, the quantization is somewhat
``trivial.'' While one can argue that $\q\restriction {\rm a}(1)$
should be faithful, there is no reason why this should be the case
for $\q$ on all of
$P$. (Indeed, if we made the latter a requirement, this would
obviate the case when the representations are \fd, since then $\q$
can never be faithful on all of $P$.) Still, we wonder if there is a
quantization which is nonzero on
$P_{(2)}\,\!$?

12. This quantization of $\T \r_+$ should be contrasted with that
given in \cite[\S 4.5]{Is}.

\ms

What makes this example work? After comparing it with our other
examples, it is clear that this polynomial quantization exists
because we can never decrease
degree in $P$ by taking Poisson brackets. Due to this we have
(\ref{ad}) as opposed to merely
\[\{P_1,P_k\} \subset P^k.\] We shall pursue this line
of investigation in a more general setting in \S 7. 

\ms

Finally, we observe that this example is symplectomorphic to $\r^2$
with the basic algebra $\sp\{p,e^{2q}\}$.

\end{subsection}


\begin{subsection}{$T^2$} 

We have just exhibited a polynomial quantization of $\T \r_+.$ But
we can do even more: Here we
exhibit a quantization of the \emph{full} Poisson algebra of the
torus. 

Consider the torus $T^2$ thought of as $\r^2/\z^2$, with symplectic
form
$$\omega=B\,dx\wedge dy.$$ We study the \ba\  $\ft$
generated by the set
\[{\cal T} = \{\sin 2\pi x,\cos 2\pi x,\sin 2\pi y,\cos 2\pi y\}.\]

\noi We already know from Proposition~\ref{prop:char} that there are
no \fd\ \ba s on the torus; thus $\ft$ is the most natural choice. 

Now $(T^2,\omega)$ is (geometrically) quantizable provided $B =
Nh$ for some nonzero integer $N$. Fix $N = 1$ and let $L$ be the
corresponding Kostant-Souriau prequantization line bundle over
$T^2$ \cite{ki}. Then the space of smooth sections
$\Gamma(L)$ can be identified with the space of ``quasi-periodic''
functions
$\varphi\in C^\infty(\r^2,\!\bc)$ satisfying
\[\varphi(x+m,\,y+n)=e^{2\pi imy}\varphi(x,\, y)\,,\quad n,\, m\in\z,\]

\noi and the prequantization Hilbert space $\h$ with the (completion
of) the set of those quasi-periodic $\varphi$  which are
$L^2$ on
$[0,\,1)\times[0,\, 1)$. The associated prequantization map
$\q:\p\to{\rm Op}\big(\Gamma(L)\big)$ (for a specific choice of
connection on $L$) is defined by 
\begin{equation}
\q(f)=-i\hbar\left[\frac{\partial f}{\partial x}\bigg(
\frac{\partial}{\partial y}-\frac{i}{\hbar}x\bigg)-\frac{\partial
f}{\partial y} \frac{\partial}{\partial x}\right]+f.
\label{eq:torusq}
\end{equation}

\noi As the torus is compact, these operators are essentially
self-adjoint on $\Gamma(L)\subset\h$.

\begin{thm}
$\q$ is a quantization of $\left(C^{\infty}(T^2),\ft\right)$.
\label{thm:nogot2}
\end{thm}

\noi {\it Proof.}  Since $\q$ is a prequantization, we need only
verify (Q4) and (Q5), (Q6) being obvious from (\ref{eq:torusq}). To
this end it is convenient to use complex notation and view
\[{\cal T}_\bc = \big\{e^{\pm 2\pi ix},e^{\pm 2\pi iy}\big\}.\] 

The analysis is simplified by applying the Weil-Brezin-Zak transform
$Z$ \cite[\S1.10]{Fo} to the above data. Define a unitary map $Z: \h
\rightarrow L^2(\r)$ by
\[(Z\phi)(x) = \int_0^1\phi(x,y)\,dy\] 

\noi with inverse
\[(Z^{-1}\psi)(x,y) = \sum_{m \in {\bf Z}}\psi(x+m)e^{-2\pi imy}.\]

\noi Under $Z$ the domain $\Gamma(L)$ maps onto the Schwartz space $\s
(\r,\!\bc)$ \cite{ki}. Setting $A_{\pm} :=Z\q(e^{\pm 2\pi ix})Z^{-1}$
and
$B_{\pm} :=Z\q(e^{\pm 2\pi iy})Z^{-1}$ we compute, as operators on
$\s (\r,\!\bc)$,
\bea (A_{\pm}\psi)(x)  & = & e^{\pm 2\pi ix}(1 \mp 2\pi ix)\psi(x)
\nonumber
\\ \rule{0ex}{4ex} (B_{\pm}\psi)(x) & = & \bigg(1 \mp 2 \pi \hbar 
\frac{d}{dx}\bigg)\psi(x \pm 1). \nonumber
\eea

\noi Then
$A_{\pm}\,\!^* = \ol{A_{\mp}}$ on the domain $\{\psi\,|\, x\psi \in
L^2(\r)\}$, and likewise $B_{\pm}\,\!^* =
\ol{B_{\mp}}$ on $\{\psi\,|\, d\psi/dx \in
L^2(\r)\}$.\footnote{\,$d\psi/dx$ is to be understood in the sense of
tempered distributions.} In fact
$\ol {A_{\pm}}$ and
$\ol {B_{\pm}}$ are normal operators.

To show that $\q(\ft)$ is an irreducible set, let us suppose that
$T$ is a bounded s.a. operator on $L^2(\bf R)$ which strongly commutes
with $\ol {A_{\pm}}$ and $\ol {B_{\pm}}$. Then $T$ must commute (in
the weak sense) with these operators on their respective
domains.\footnote{\,Here and in what follows we use the fact that a
bounded operator weakly commutes with an (unbounded) normal operator
iff they strongly commute.} Consequently $T$ commutes with both
\begin{equation}
{\ol{A_-}}\,{\ol{A_+}} = I + 4 \pi^2 x^2
\label{eq:classidT2}
\end{equation}

\noi  on the domain $\{\psi\,|\,x^2\psi \in L^2(\r)\}$, and
\[{\ol{B_-}}\,{\ol{B_+}} = I - 4\pi^2 \hbar^2 \frac{d^2}{dx^2}\]

\noi on $\{\psi\,|\,d^2\psi/dx^2 \in L^2(\r)\}$. {}From these
equations we see that
$T$ commutes, and therefore strongly commutes, with the closures of
two of the three generators of the metaplectic representation
(\ref{eq:p2}) of sp$(2,\r)$ on
$\s (\r,\!\bc)$.

Suppose that $\mu$ denotes the metaplectic representation of the
metaplectic group Mp(2,$\r$) on
$L^2(\r)$. We have in effect just established that $T$ commutes with
the one-parameter groups
$\exp\!\big(is\,{\ol{x^2}}\big)$ and
$\exp\!\big(\!-\!it\hbar^2\,{\ol{d^2/dx^2}}\big)$. Now classically the
exponentials
$\exp(sx^2)$ and
$\exp(ty^2)$ generate Sp$(2,\r)$ \cite[\S 4]{GS}. As Mp$(2,\r) \ra
{\rm Sp}(2,\r)$ is a double covering, the corresponding exponentials
in Mp(2,$\r$) generate a neighborhood of the identity in the
metaplectic group.  Since $\mu\big[\exp(sx^2)\big] =
\exp\!\big(is\,{\ol{x^2}}\big)$ and
$\mu\big[\exp(ty^2)\big] =
\exp\!\big(\!-\!it\hbar^2\,{\ol{d^2/dx^2}}\big)$, it follows that $T$
commutes with $\mu({\cal M})$ for all ${\cal M}$ in a neighborhood of
the identity in Mp(2,$\r$) and hence, as this group is connected, for
all
${\cal M} \in$ Mp$(2,{\bf R})$.

Although the metaplectic representation $\mu$ is reducible, the
subrepresentations
$\mu_e$ and $\mu_o$ on each invariant summand of $L^2({\bf R})  =
L^2_{e}({\bf R})
\oplus L^2_{o}({\bf R})$ of even and odd functions are irreducible
\cite[\S 4.4]{Fo}. Writing $T = P_eT + P_oT$, where $P_e$ and $P_o$
are the even and odd projectors, one has
\begin{equation} [P_eT,\mu({\cal M})] = 0 \label{eq:com}
\end{equation}

\noi for any ${\cal M} \in \mbox{Mp(2,}{\bf R})$. It then follows from
the irreducibility of the subrepresentation $\mu_e$ that
$P_eT = k_eP_e + RP_o$ for some constant $k_e$ and some operator
$R:L^2_{o}({\bf R}) \ra L^2_{e}({\bf R})$. Substituting this
expression into (\ref{eq:com}) yields $[RP_o,\mu({\cal M})] = 0$, and
Schur's Lemma then implies that $R$ is either an isomorphism or is
zero. But $R$ cannot be an isomorphism as the representations $\mu_e$
and $\mu_o$ are inequivalent \cite[Thm.~4.56]{Fo}. (Recall that two
unitary representations are similar iff they are unitarily
equivalent.) Thus
$P_eT = k_eP_e$. Similarly
$P_oT = k_oP_o,$ whence $T = k_eP_e + k_oP_o.$ 

But now a short calculation shows that $T$ commutes with 
\[{\ol {A_+}}-{\ol{A_-}} = 2i(\sin 2\pi x - 2\pi x\cos 2\pi x)\] 

\noi only if $k_e = k_o$. Thus $T$ is a multiple of the identity, and
so $\{A_{\pm},B_{\pm}\}$ is an irreducible set, as was to be shown.
Thus in particular (Q4) is satisfied.

For (Q5), we claim that the linear span of the Hermite functions form
a dense set of separately analytic vectors for the e.s.a.\ components
of
$\{A_{\pm},B_{\pm}\}$. {}From the expression above for $A_{\pm}$, it
is clear that a vector will be analytic for the e.s.a.\ components of
$A_{\pm}$ iff it is analytic for multiplication by $x$. But it is well
known that the Hermite functions are analytic for this latter
operator. The corresponding result for
$B_{\pm}$ is obtained directly from this by means of the Fourier
transform.
\endproof

\ms

\noi \emph{Remark}. 13. The proof also works for $N =-1$ but breaks
down when
$|N| \neq 1$ \cite{Go}. It is not known to what extent this
theorem will remain valid in general (but see \S 7). As a consequence
the classical limit is unclear; to compute it, one needs to study
how the torus quantization behaves for large values of the
quantum number $N$. But for
$N > 1$, the {\it pre\/}quantizations with Chern
class $N$ may not be actual quantizations. If they are not, then one
must construct a series of quantizations $\q_1,\ldots,\q_N,\ldots$
with $\q_1= \q$ and see what happens to $\q_N$ as $N$ grows. Without
these ``interpolating quantizations,'' the classical limit of $\q$
cannot be determined. 

\ms 

This full quantization has several remarkable features. (See
\cite{Go,Ve} for detailed discussions). First, in
previous examples the irreducibility requirement typically led to
\vn\ rules. But for
$T^2$ both
$\q(f^2)$ and $\q(f)^2$ are completely determined
for any observable $f$ by the simple fact that
$\q$ is a prequantization; irreducibility is irrelevant.
Moreover, one sees from
(\ref{eq:torusq}) that $\q(f^2)$ is a first order
differential operator whereas $\q(f)^2$ is of second order,
 indicating that this
quantization will not respect the classical multiplicative
structure at all.

This is particularly evident when one considers the classical
identity 
\linebreak
$\cos^2 2\pi x + \sin^2 2\pi x = 1,$ as emphasized by
\cite{Ve}. In view of (\ref{eq:classidT2})
\begin{equation}
\left[\q(\cos 2\pi x)\right]^2 + \left[\q(\sin 2\pi
x)\right]^2 = I + 4\pi^2 x^2,
\label{eq:quantidT2}
\end{equation}

\noi which bears scant resemblance to 
\[\q(\cos^2 2\pi x) + \q(\sin^2 2\pi x) = I.\]

\noi So the torus quantization dramatically violates Souriau's
requirement that ``the quantum spectrum of commuting observables
should be concentrated on their classical range'' \cite{Z}. As
reflected by (\ref{eq:quantidT2}), the bounded observables $\cos 2\pi
x$ and $\sin 2\pi x$ quantize to unbounded operators. While this
may be seen as a flaw of the quantization, it cannot be helped: A
theorem of Avez states that when the phase space $M$ is compact, 
the only possible prequantization of
$C^{\infty}(M)$ by \emph{bounded} operators is $f \mapsto {\bar f}I$,
where $\bar f$ is the mean value $f$ \cite{av1}. 
If the torus is to be fully (and nontrivially) quantized,  the
representation space must thus be infinite-dimensional, whence a
certain ``amount'' of unboundedness must ensue. So in this regard,
the torus is not really behaving badly; there is a trade-off involved
here.

Finally, the salient feature of this example
is that the \ba\ $\ft$
is \id. This also did not happen in any of our other examples. As a
consequence the irreducibility requirement on
$T^2$ is substantially weaker than the corresponding requirements
on either $\r^{2n}$, $S^2$, or $\T\! S^1$, and is likely the
underlying reason why $\q$ provides a full quantization of
$\big(C^{\infty}(T^2),\ft\big)$ 

\end{subsection}

\end{section}


\begin{section}{No-Go Theorems}

Our treatment of the examples in \S 5 relied heavily on an
intimate knowledge of the representations of the relevant \ba s, and
involved detailed calculations. Here we present some
general results on the occurrence of obstructions. To accomplish this,
we focus on the Lie and Poisson structures of \ba s and the
polynomial algebras they generate; necessarily, the
representations of these objects now play a more subdued
role. Background on Poisson algebras is given in
\cite{At,Gr1,Gr2}.

The first key result appeared in 1974 and is due to Avez
\cite{av1,av2}. Recall that the mean value of $f \in \ci (M)$ is
\[\bar f = \frac{1}{\mbox{vol}(M)}\,\int_M f\,\omega^n.\] 
\begin{thm}
The only full prequantization of a compact symplectic manifold by
\linebreak bound\-ed operators 
is given by $f \mapsto \bar fI$. 
\label{thm:avez}
\end{thm}

Thus there can be no nontrivial \fd\ full prequantizations of a
compact phase space. In the noncompact case, there is the following
complementary  result due to Doebner and Melsheimer \cite{d-m}. 
\begin{prop}
\label{thm:d-m}
A nonzero \id\ representation of a noncompact \fd\ Lie algebra by
skew-symmetric operators contains at least one unbounded operator.
\end{prop}

Combining these two results, we see that an \id\ quantization
will necessarily involve unbounded operators. Whereas
Theorem~\ref{thm:avez} uses the Poisson structure on
$\ci (M)$, Proposition~\ref{thm:d-m} is purely representation
theoretic. We shall encounter this dichotomy again in \S 6.3. 

The next
advance was made by Ginzburg and Montgomery
\cite{GM}, who generalized Avez's theorem to noncompact $M$.
Let $C^\infty_c(M)$ denote the Poisson algebra  of compactly
supported smooth functions on $M.$ 
\begin{thm} 
\label{thm:gm} There is no nontrivial finite-dimensional Lie
representation of $C^{\infty}_c(M).$
\end{thm}

We do not give the proof, as it is similar to that of
Theorem~\ref{thm:gg} following. Since a prequantization is simply a
special type of Lie representation, Theorems~\ref{thm:avez}  and
\ref{thm:gm} yield the no-go result:
\begin{cor} There exists no nontrivial finite-dimensional full
prequantization of any symplectic manifold $M.$
\label{cor:gmp}
\end{cor} 

Inspired by this work, we generalize both
Theorem~\ref{thm:gm} and Corollary~\ref{cor:gmp} to polynomial
quantizations. Let
$\fb$ be a basic algebra of observables
and
$P(\fb)$ the \pa\ of polynomials generated by $\fb$. \emph{Throughout
this section we assume that $\fb$ is \fd}.
We break the analysis up into four cases, depending upon whether 
$\fb$, or equivalently $M$, is compact and its representations are
\fd. It turns out that we are able to obtain obstructions to
quantizing
$(P(\fb),\fb)$ in three of these cases. And in the remaining case
(viz. when $\fb$ is noncompact and the representation space is \id),
there is no universal obstruction. In this gross sense,
then, we have completely solved the
\gr -\vh\ problem for polynomial quantizations.


\subsection{$M$ Compact, Finite-dimensional
Representations}

The main result is:
\begin{thm}
\label{thm:gg} 
Let $\fb$ be a \fd\ \ba\ on a compact symplectic manifold $M$. There
exists no nontrivial finite-dimensional Lie representation of
$P(\fb)$.
\end{thm}

We begin with a purely
algebraic lemma, whose proof is given in \cite{GGG}. 
\begin{lem}
\label{lem:gra} If $L$ is a finite-codimensional Lie ideal of an \id\
Poisson algebra $\cal P$ with identity, then either $L$ contains the
commutator ideal
$\{ \cal P,\cal P\}$ or there is a maximal
finite-codimensional  associative ideal
$J$ of $\,\cal P$ such that $\{ \cal P,\cal P\} \subseteq J$.
\end{lem}

\noi \emph{Proof of Theorem \ref{thm:gg}.} Suppose that $\q$ were a
Lie representation of $P(\fb)$ on some \fd\ vector space. Then $L =
\ker
\q$ is a \fc\ Lie ideal of $P(\fb).$ We will show that $L$ has 
codimension at most 1, whence the representation is trivial. We
accomplish this in two steps, by showing that:

\begin{description}
\item \rule{0mm}{0mm}
\begin{enumerate}
\vspace{-4ex}
\item[(a)] The derived ideal $\{P(\fb),P(\fb)\}$ has codimension 1 in
$P(\fb),$ and
\vskip 6pt
\item[(b)]  $L \supseteq \{P(\fb),P(\fb)\}$.
\end{enumerate}
\end{description}

Let $A(\fb)$ denote the Lie
ideal of polynomials of zero mean. The decomposition $f
\mapsto \bar f + (f -
\bar f)$ gives 
$P(\fb) = \r \oplus A(\fb)$. Thus, if we prove that
$\{P(\fb),P(\fb)\} 
\linebreak
= A(\fb)$, (a) will follow.

Using (\ref{eq:identity}) along with Stokes' Theorem, we immediately
see that
$\{P(\fb),P(\fb)\}
\linebreak
\subseteq
A(\fb)$. To show the reverse inclusion, let
$\{b_1,\ldots,b_N\}$ be a basis for
$\fb$, so that
\[\{b_i,b_j\} = \sum_{k=1}^N c^k_{ij}b_k\]

\noi for some constants $c^k_{ij}$. Following Avez \cite{av2},
define the ``symplectic Laplacian''
\[\mathnormal{\Delta}f = - \sum_{i = 1}^N \,\{b_i,\{b_i,f\}\}.\]

\noi It is clear from these two expressions and the Leibniz rule that
the linear operator
$\mathnormal{\Delta}$ maps
$P^k(\fb)$ into $A^k(\fb).$ Furthermore, taking into account the
transitivity of $\fb$, we can apply
\cite[Prop.~1(4)]{av2} to conclude that
$\mathnormal{\Delta}f = 0$ only if $f$ is constant. Thus for each $k
\geq 0$, the decomposition $P^k(\fb) = \r \oplus
A^k(\fb)$ implies
$\mathnormal{\Delta}(P^k(\fb)) = A^k(\fb).$ It follows that
$A(\fb) \subseteq
\{P(\fb),P(\fb)\}.$

If (b) does not hold, then by Lemma \ref{lem:gra} there must be a
proper associative ideal $J$ in $P(\fb)$ with $\{P(\fb),P(\fb)\} 
\subseteq J.$ Since
$\{P(\fb),P(\fb)\}= A(\fb)$ has codimension 1, $A(\fb) = J.$ This is,
however, impossible, since $f^2$ has zero mean only if $f=0. $
\endproof

\ms

\begin{cor}
\label{cor:pq} Let $\fb$ be a \fd\ \ba\ on a compact
symplectic manifold $M$. There exists no nontrivial
finite-dimensional prequantization of
$P(\fb)$. In particular, there exists no nontrivial \fd\
quantization of $(P(\fb),\fb)$.
\end{cor}

Although
not surprising on mathematical grounds, since
$P(\fb)$ is ``large,'' these corollaries do have physical
import, as one expects the quantization of a compact phase
space to yield a
\emph{finite-}dimensional Hilbert space.


\subsection{$M$ Compact, Infinite-dimensional
Representations}

We reduce this to the previous case as follows. Suppose that $\q$
were a quantization of
$(P(\fb),\fb)$ on a Hilbert space. By conditions (Q3) and (Q5),
$\q(\fb)$ can be exponentiated to a unitary representation of the
simply connected Lie group $B$ with \la\ $\fb$ (recall
that $\fb$ is assumed \fd) which, according to (Q4), is irreducible.
Since $M$ is compact, $B$ is compact. The representation
space must thus be \fd, and so Corollary \ref{cor:pq} applies. This
proves

\begin{thm}
\label{thm:ggg}
 Let $\fb$ be a \fd\ \ba\ on a compact symplectic manifold $M$. There
exists no nontrivial quantization of $(P(\fb),\fb)$.
\end{thm}

Thus, there is an obstruction to polynomially quantizing a
compact symplectic manifold \emph{regardless} of the dimensionality
of the representation.


\subsection{$M$ Noncompact, Finite-dimensional
Representations}

Now suppose that $M$ is noncompact. On
physical grounds  one expects a quantization of $M$, if it
exists, to be infinite-dimensional. This is what we
rigorously prove here.

Already on the basis of representation theory, one can see that it
will be difficult to obtain \fd\ quantizations of noncompact \ba s.
For instance, it is known that a Lie algebra admits a
nontrivial \fd\ irreducible representation by skew-symmetric operators
iff its Levi factor contains a nontrivial compact ideal
\cite[Prop.~8.7.3]{b-r}. Thus in particular a solvable algebra has no
nontrivial
\fd\ irreducible representations. We now prove that a \emph{basic}
algebra cannot admit any faithful \fd\ representations at all,
irreducible or not.
\begin{thm} Let $\fb$ be a \fd\ \ba\ on a noncompact symplectic
manifold. Then $\fb$ has no faithful \fd\ representations by
symmetric operators.
\label{thm:gg2}
\end{thm}

\noindent {\it Proof.} We argue by contradiction. Suppose
there exists a representation $\varrho$ of $\fb$ on some $\bc^k$. As
$\varrho(\fb)$ consists of hermitian matrices, $\varrho$ is 
completely reducible. Since by assumption $\varrho$ is faithful, one
deduces from
\cite[Theorem 3.16.3]{V} that $\fb$ is reductive.
By the comment following the the proof of
Proposition~\ref{prop:bs}, $\fb$ must then be semisimple. 

Since $M$ is noncompact, so is the simply connected covering group
$B$ of the semi\-simple algebra $\fb$. Now consider a unitary
representation
$U$ of
$B$ on
$\bc^k$. Decompose $B$ into a product
$B_1 \times \cdots \times B_K$ of simple groups. Then (at least)
one of these, say
$B_1$, must be noncompact. But it is well-known that a
connected, simple, noncompact Lie group has no nontrivial \fd\ unitary
representations \cite[Theorem. 8.1.2]{b-r}. Thus 
$U(b) = I$ for all $b \in B_1$. Since every finite-dimensional
representation
$\varrho$ of
$\fb$ by symmetric operators is a derived representation of some
unitary representation $U$ of $B$, it follows that
$\varrho\restriction {\fb}_1 = 0$, and so $\varrho$ cannot be
faithful.
\endproof

\ms

Since every quantization of $(\oo,\fb)$ must be faithful
on $\fb$, we conclude that \emph{there is no nontrivial \fd\
quantization of $(\oo,\fb)$ on a noncompact symplectic
manifold,} where $\oo$ is \emph{any} (unital) \la\ containing $\fb.$
Combining this with Corollary~\ref{cor:pq} we can now
assert---roughly speaking---that no symplectic manifold with a (\fd)
\ba\ has a \fd\ quantization.


\subsection{$M$ Noncompact, Infinite-dimensional
Representations}

So far we have encountered obstructions in every
instance. The present case is the exception: We know from \S 5.4 that
there exists a polynomial quantization of $\T \r_+$ with  the
basic algebra a(1). 

The behavior exhibited by this example is not characteristic of
solvable algebras such as a(1), since e(2) for the
cylinder is also solvable yet exhibits an obstruction. Likewise, the
Heisenberg algebra is nilpotent and is obstructed as well. 


\subsection{Discussion and Further Results}

Theorem \ref{thm:ggg} asserts that the polynomial algebra $P(\fb)$
generated by any finite-dimen\-sion\-al basic algebra $\fb$ on a
compact symplectic manifold cannot be consistently quantized. As the
torus illustrates, this need not be true if $\fb$ is
allowed to be infinite-dimen\-sion\-al. Similarly Theorem
\ref{thm:gg} and Corollary
\ref{cor:pq} can fail when the  representation space is allowed to be
\id: as is well-known, full prequantizations exist provided
$\omega/h$ is integral. Thus Corollary \ref{cor:pq} and Theorem
\ref{thm:ggg} are the optimal no-go results for compact phase spaces.

When $M$ is compact, Proposition~\ref{prop:char} enables us to identify
$P(\fb)$ with the Poisson algebra of polynomials on
$\fb^*$ restricted to the coadjoint orbit $M$. In particular, we can
take
$M=S^2
\subset
\mbox{su(2)}^*$,
$\fb$ the space of spherical harmonics of degree one ($\fb \cong
\mbox{su}(2)$), and
$P(\fb)$ the space of all spherical harmonics.  Thus
Theorem~\ref{thm:nogos2} follows immediately from Theorem
\ref{thm:ggg}. A similar analysis applies to
$\bc P^n
\subset
\mbox{su}(n+1)^*$.

Our results in the compact case lean heavily on the algebraic
structure of $P(\fb)$, and in particular on the property that
$\{P(\fb),P(\fb)\}$ has codimension 1 in $P(\fb)$. When $M$
is noncompact, $\codim
\{P(\fb),P(\fb)\}$ is not fixed; it takes on the values $0$, $1$,
and even $\infty$ in examples. Thus the Poisson theoretic techniques
that worked for compact phase spaces will not apply to noncompact
ones. This partly explains why Theorem~\ref{thm:gg2} is a
representation theoretic result. Furthermore, this theorem
hinges on the fact that
$\fb$, being noncompact and semisimple, cannot have faithful
\fd\ representations by Hermitian matrices. But when $M$ is 
compact, $\fb$ is compact semisimple, and these algebras \emph{do}
have such representations. Thus the compact and noncompact cases
require entirely different approaches. 

It is useful to keep track of which hypotheses the five theorems in
this section require. They all use (Q1), and
Theorems~\ref{thm:gm} and \ref{thm:gg} require only
this. Theorem~\ref{thm:avez} needs (Q2) as well.
Theorem~\ref{thm:ggg} uses also (Q3)--(Q5), and lastly
Theorem~\ref{thm:gg2} assumes in addition only (Q6). 
We do not know if a no-go theorem can be proven in the noncompact,
\fd\ case without the faithfulness assumption (Q6). Irreducibility
was only used in the proof of Theorem~\ref{thm:ggg}; in the other
cases the \fd ity assumption forced the representation to be
``small.''

\ms

We are thus left with trying to understand the noncompact,
infinite-dimen\-sion\-al case, which is naturally
the most difficult one. Here one has little control over
either the types of \ba s that can appear (in examples they range from
solvable to semisimple; compare Proposition~\ref{prop:bs}), the
structure of the polynomial algebras they generate (cf. the above),
or their representations. Thus one should try a different tack.
Following the lead of Joseph \cite{Jo} (cf. \S 5.1), let us try
to compare the algebraic structures of Poisson algebras on the one
hand with associative algebras of operators with the commutator
bracket on the other. Grabowski has adopted this approach, and has
produced the following ``algebraic'' no-go theorem, which is proved
in \cite{GGra}.
\begin{thm}
\label{thm:algnogo}
Let $\cal P$ be a unital Poisson subalgebra of $\,\ci (M).$ If as a
\la\
$\cal P$ is not commutative, it cannot be realized as an associative
algebra with the commu\-ta\-tor bracket.
\end{thm}

Apply this result to polynomial quantizations. Take $\cal P =
P(\fb)$, and let $\varrho: \fb \ra {\rm Op}(D)$ be a representation of
$\fb.$ Define $\cal A \subset {\rm Op}(D)$ to be the
associative algebra generated by $\{\varrho(f)\,|\,f \in \fb\}$
together with $I$ (if $1 \not \in \fb$). Suppose $\q$ is a
quantization of
$P(\fb)$ which is valued in
$\cal A$ and extends
$\varrho$. If it can be shown that any such $\q$ must be a \la\
isomorphism of
$P(\fb)$ onto $\cal A$, then the algebraic no-go theorem
will yield a contradiction.

To see how this works in practice, let us once again look at the
Heisenberg algebra on $\r^2.$ \emph{Supposing} that $\q\big(P(\rm
h(2))\big) \subset \cal A,$ we can then use the argument
of \cite{GGra} to show that $\q$ is surjective. We shall prove
inductively that
\[\q(q^kp^l)=X^kY^l+\sum_{k'+l'<k+l}a^{kl}_{k'l'}X^{k'}Y^{l'}\]

\noi for some constants  $a^{kl}_{k'l'}$,  where  $X=\q(q),\ 
Y=\q(p)$. Indeed,
\begin{eqnarray*} [\q(q^kp^l),Y] & = & -i\hbar \q(\{q^kp^l,p\})
\hspace{1ex} =  \hspace{1ex} i\hbar
k\q(q^{k-1}p^l) \\ \rule{0ex}{3ex}
& =& i \hbar kX^{k-1}Y^l+ {\it \ lower\ degree\ terms},
\end{eqnarray*}

\noi where we have used the inductive
assumption. Similarly
\[ [\q(q^kp^l),X]=-i\hbar lX^kY^{l-1}+{\it\ lower\ degree\ terms}.\]

\noi Due  to  $ad_X\circ  ad_Y=ad_Y\circ  ad_X$,  we   can   find
$F^{kl}=X^kY^l+ {\it\ lower\ degree\ terms}$,  which  has  the 
same commutators with $X$ and $Y$ as $\q(q^kp^l)$. Hence, \emph{if}
$\q$ is ``algebraically irreducible'' in the sense
that the the only elements of $\cal A$ which commute with
$\q(\rm h(2))$ are multiples of the identity,
then $\q(q^kp^l)$ differs from $F^{kl}$
by  a  constant, and that proves the inductive step.

Now it is easy to see that every
nontrivial Lie ideal of
$P({\rm h}(2))= \r[q,p]$ intersects $P^1$. In particular if $\ker
\q \neq \{0\}$, then we contradict either (Q1) or (Q2). Thus
$\q$ must be injective, and so we have an algebraic obstruction to
quantizing
$\big(P(\rm h(2)),\rm h(2)\big)$.

\ms 
This argument can likely be extended to any nilpotent \ba\
\cite{GGra}.

As highlighted above, there are two difficulties in
correlating the algebraic approach with our previous
considerations. The first is that the connection between the
``algebraic irreducibility'' used here and the ``analytic
irreducibility'' of \S 4 is unclear, due to functional analytic
subtleties. The second is that there is no
\emph{a priori} reason why
$\q\big(P(\rm h(2))\big) \subset \cal A.$ This requirement is
reminiscent of a \vn\ rule. Perhaps the analytic irreducibility
condition (Q4) can be used to establish this inclusion; this is
actually the case for the Heisenberg algebra as shown by
Proposition~\ref{prop:fix} and Lemma~\ref{lem:morevnrs}.

Regardless, it appears that this algebraic approach holds promise; at
least it enables us to partially suppress the
representational aspects over which we have little control.

\end{section}


\begin{section}{Speculations}

In view of the theorems in the previous section, obstructions to
quantization are guaranteed to exist except when the phase
space is noncompact and the representations under consideration are
\id. Three of our examples fall into this category: $\r^{2n}$, $\T\!
S^1$, and $\T \r_+.$ The first two exhibit obstructions, while the
last does not. Comparing the behavior of these examples, as well as
that of $S^2$, which is also obstructed, we attempt to extract the key
features which govern the appearance of obstructions to a polynomial
quantization.

Of course, any conclusions that we can draw at this point are
necessarily tentative, due to the paucity of examples against
which to test them. There are also various aspects of these 
examples that still are  not completely understood. Nonetheless, some
interesting observations can be made, which may prove helpful in
subsequent investigations. 

A detailed look at the derivations of the \vn\ rules
for $\r^{2n}$, $\T\! S^1$, and $S^2$, and how they engender
obstructions, shows that the controlling factor is apparently that
one can decrease degree \emph{in} $P(\fb)$ by taking Poisson brackets.
This is particularly evident in the classical \pb\ relations
(\ref{eq:s2pbr}) and (\ref{eq:s2pbr2}), and (\ref{eq:T*s1pbr}),  which
led to the contradictions for $S^2$ and
$\T\! S^1$, respectively. The situation for $\r^2$ is subtler,
but one can spot this phenomenon in the proofs of
Proposition~\ref{prop:fix} and Lemma~\ref{lem:morevnrs}. The analysis
in
\S 5.4 shows that it is
\emph{not} possible to decrease degree in $P(\fb)$ by taking \pb s on
$\T \r_+.$

There are two---and only two---circumstances under which taking \pb s
in $P(\fb)$ can decrease degree:\footnote{\,\emph{A priori,} a third
circumstance would be if $1 \in \fb$. Using the minimality condition
(B4), it is not difficult to prove that then $1 \in \{\fb,\fb\}$, so
this is actually a subcase of (D1).} 

\begin{description}
\item \rule{0mm}{0mm}
\begin{enumerate}
\item[(D1)] $1 \in \{P(\fb),P(\fb)\}$, and
\vskip 6pt
\item[(D2)] $P(\fb)$ is not free as an associative algebra.
\end{enumerate}
\end{description}

\noi According to the discussion at the end of \S 3, (D2) is
equivalent to the existence of Casimirs in the symmetric algebra of
$\fb$. Thus (D2) holds whenever $\fb$ is semisimple, and in
particular when it is compact (cf. the comment following the proof of
Proposition~\ref{prop:bs}). At the other extreme, when
$\fb$ is nilpotent, (D1) holds. Indeed, a nilpotent algebra has a
center, and (B3) implies that this center consists of constants. An
examination of the descending central series for
$\fb$ then shows that $1 \in \{\fb,\fb\}$. In the examples,
$\r^{2n}$ satisfies (D1) but not (D2),
$S^2$ satisfies (D2) by virtue of (\ref{eq:s2casimir}) but not
(D1), and $\T\! S^1$ satisfies both
because of
\[1= \cos^2 \theta + \sin^2 \theta = \textstyle \frac{1}{2}
\big\{\{\ell^2,
\sin \theta\},\sin \theta\big\} + \textstyle\frac{1}{2} 
\big\{\{\ell^2,
\cos \theta\},\cos \theta\big\}.\] 

\noi On the other hand, $\T \r_+$ satisfies neither condition.

On the basis of this ``anecdotal'' evidence, we
propose that a general Groene\-wold-Van~Hove theorem takes the form:
\begin{conj} Let $M$ be a symplectic manifold with  a \fd\ basic
algebra $\fb$. Suppose that the polynomial algebra $P(\fb)$
satisfies either {\rm (D1)} or {\rm (D2)}. Then there is no
nontrivial quantization of $(P(\fb),\fb)$.
\label{conj:gvh}
\end{conj}

Indeed, is possible to directly verify this
conjecture under certain circumstances.
\begin{thm} Conjecture {\rm \ref{conj:gvh}} is valid 
when either $M$ is compact or the representation space is \fd.
\label{thm:gvhpfcpt}
\end{thm}

\noi {\it Proof.} According to
Proposition~\ref{prop:bs}, when $M$ is compact $\fb$ is compact. 
Just as in \S 6.2 we may then use (Q3)--(Q5) to reduce the case of
infinite-dimen\-sion\-al representations to that of \fd\ ones.  Thus
it suffices to prove the theorem for the case when $\q$ is a
quantization of
$P(\fb)$ on a \fd\ Hilbert space, whence $L = \ker \q$ has finite
codimension in $P(\fb)$. 

Arguing as in the proof of Theorem~\ref{thm:gg2}, we have from
(Q6) that $\fb$ is semisimple. Furthermore, (Q4) and
\cite[Prop.~8.7.3]{b-r} imply that $\fb$ contains a nonzero compact
ideal
$\mathfrak a$.  

We apply
Lemma~\ref{lem:gra} to $L$. First suppose that
$\{P(\fb),P(\fb)\} \subseteq L$.
Then semisimplicity gives
$\fb = \{\fb,\fb\} \subset L$, and so  $\q\restriction\fb = 0$, which
contradicts (Q6). 

Thus there must exist a maximal \fc\ associative ideal $J$ in
$P(\fb)$ with $\{P(\fb),P(\fb)\} \subseteq J.$ If (D1) holds, then $1
\in J$, which cannot be as
$J$ is proper. Now suppose (D2) holds (as in fact it must, as $\fb$
is semisimple), so that there is a Casimir
$C \in S(\fb)$. If
$\rho$ is the projection $S(\fb) \ra P(\fb)$, then $K =
\rho^{-1}(J)$ is a maximal \fc\ associative ideal in $S(\fb)$ with
$\{S(\fb),S(\fb)\} \subseteq K.$ Since $\fb =
\{\fb,\fb\} \subset \{S(\fb),S(\fb)\} \subseteq K$, and since $1 \not
\in K$ (as $K$ is proper), it follows that
$K$ is the associative ideal generated by $\fb.$ (Actually, this
shows that $S(\fb) = \r \oplus K.$)

Since $C$ is a Casimir, transitivity implies that $\rho(C) = c$ for
some constant $c.$  Let $C_2$ be the quadratic Casimir of
the ideal $\fa \subseteq \fb$; then $C_2$ is also a Casimir for $\fb$
and, since
$\mathfrak a$ is compact,
$C_2$ is negative-definite:
$\rho(C_2) = c_2 < 0.$ Now choose a constant $k$ such that
$\rho(C + kC_2) = c + kc_2 < 0$.
By the definition of a Casimir and the above remarks $C + kC_2 \in K.$
But then
$(C + kC_2) - (c + kc_2) \not \in K$, which is a contradiction since
$(C + kC_2) - (c + kc_2) \in \ker \rho \subset K.$
\endproof

\ms

While similar to the proof of Theorem~\ref{thm:gg}, this proof
has a key advantage: It does not require us to know the
detailed structure of the commutator ideal (which we do not, when
$\fb$ is noncompact).

Thus Conjecture~\ref{conj:gvh} is consistent with the results of \S
6. Furthermore, the hypotheses of Conjecture~\ref{conj:gvh} are
certainly
\emph{necessary}. 
\begin{thm}
Suppose that the polynomial algebra $P(\fb)$
satisfies neither condition {\rm (D1)} nor {\rm (D2)}. Then any
nontrivial quantization of $\fb$ extends to a 
quantization of $(P(\fb),\fb)$.
\label{thm:conconj}
\end{thm}

\noi {\it Proof.} 
For if $P(\fb)$
satisfies neither of these conditions, then the notion of
homogeneous polynomial is well-defined and it is not
possible to lower degree in $P(\fb)$ by taking \pb s. Just as in
\S 5.4,  $P_{(2)}(\fb)$
is then an ideal in $P(\fb)$, and $P(\fb) = P^1(\fb) \ltimes
P_{(2)}(\fb)$. Let $\varrho$ be the assumed
representation of
$\fb$;  this
extends to a representation of $P^1(\fb)$. Then $\q = \varrho \oplus
0$ is the required quantization of $(P(\fb),\fb).$
\endproof

\ms

Lastly, we observe that the \fd ity assumption on $\fb$ in
Conjecture~\ref{conj:gvh} is necessary as well: The symmetric algebra
$S(\ft)$ on $T^2$ certainly contains Casimirs, but violates
the conjecture.

\ms

 Of our five examples, the torus is clearly much different
than the others. It is not a Hamiltonian homogeneous space, and the
\ba\ $\ft$ is infinite-dimensional. Because of this, the
irreducibility requirement (Q4) loses much of its force -- so much
so that it precludes the existence of an
obstruction. So it seems equally reasonable to propose
\begin{conj} Let $M$ be a symplectic manifold and
$\fb$ a basic algebra with $P^1(\fb)$ dense in
$C^{\infty}(M)$.\footnote{\,We use $P^1(\fb)$ here to ensure that 1
is present: On the torus, $\fb$ consists only of trigonometric
polynomials of mean zero, whereas $P^1(\fb)$ comprises all
trigonometric polynomials.} Then there exists a nontrivial
quantization of
$(\ci (M),\fb)$.
\label{conj:go}
\end{conj}

A necessary condition for $\q$ to be a full quantization of
$(\ci (M),\fb)$ is that $\q$ represent $\ci (M)$ itself irreducibly.
It turns out \cite{ch3,Tu95} that this is so for all Kostant-Souriau
prequantizations\footnote{\,However, there are
other prequantizations which do not represent $\ci (M)$ irreducibly;
for instance, the prequantization of Avez \cite{av,Ch2}.}; thus it is
natural to consider the case when $M$ is prequantizable in this sense.
In fact, in this context \cite{Tu95} gives even more:
\begin{prop} Let $M\!$ be an integral symplectic manifold, $\!L$ a 
Kostant-Souriau prequantization line bundle over $M$ and
$\q_L$ the corresponding prequantization map. Let
$\fb$ be a basic algebra with
$P^1(\fb)$ dense in $C^{\infty}(M)$. Then $\q_L$ represents $\fb$
irreducibly on the domain consisting of compactly supported
sections of $L$.
\label{prop:gijs}
\end{prop}

Set $D_c =
\Gamma(L)_c$, the compactly supported sections of $L$. By construction
$\q_L:\ci (M) \ra {\rm Op}(D_c)$ satisfies (Q1)--(Q3) and (Q6). This
proposition states that $\q_L$ satisfies (Q4) as well. Thus to obtain
a full quantization it remains to verify (Q5)---perhaps on
some appropriately chosen coextensive domain $D$; unfortunately, it
does not seem possible to do this except in specific instances. A
first test would be to understand what happens for
$\big(\ci (T^2),\ft\big)$ with $|N| \neq 1$. In any event,
Proposition~\ref{prop:gijs} does provide a certain amount of support
for Conjecture~\ref{conj:go}. 

The ``gray area'' between these two conjectures consists of symplectic
manifolds with \ba s $\fb$ for which $P^1(\fb)$ is
infinite-dimensional, yet not dense in 
$C^{\infty}(M)$. Maybe the infinite-dimensionality of
$\fb$ alone is enough to guarantee the existence of a full
quantization?

\ms

Completing the proof of Conjecture~\ref{conj:gvh}---that is, when $M$
is noncompact and the quantizations are \id ---seems to be a
difficult problem. Perhaps the ``algebraic approach'' sketched at the
end of \S 6 will prove useful; it already appears promising when
$\fb$ is nilpotent \cite{GGra}. It will likely be necessary to work
through a few more examples of
\gr-\vh\ obstructions before one is able to gain sufficient insight
into this problem. One example worth studying are the various
coadjoint orbits for sp(2$n,\r$).
As well, it would be useful to consider basic algebras of a more
general type than the ones we have encountered thus far (which were
all either solvable or semisimple). We have also restricted
consideration to polynomial subalgebras to a large extent, but there
are other subalgebras $\oo$ which are of interest (e.g., on
$\r^{2n}$, those functions which are constant outside some compact set
\cite{Ch2}).

A negative answer to the conjecture might indicate that one should
strength\-en the conditions defining a basic algebra by, e.g.,
replacing (B3) by (C2) as discussed in \S 3 (although this
specific change would eliminate a(1) on $\T\r_+$ from the ranks of
\ba s.) One could also modify the axioms for a quantization, for
instance by adopting Souriau's requirement that classical observables
with bounded spectra should quantize to operators with bounded
spectra.  Or, if the conjecture still seems undecidable, perhaps one
should abandon the definition of a quantization map solely in terms
of basic algebras and consider an alternative. However, the two other
ways to define a quantization map listed previously suffer from
serious flaws. If one imposes
\vn\ rules at the outset, then one tends to run into difficulties
rather quickly---especially if one tries to enforce the rules on all
of $\ci (M)$ and not some basic algebra thereof---as was shown in
\S 5.1. Furthermore, it is unclear what form \vn\ rules should take in
general, as is illustrated by the unintuitive rules
(\ref{eq:sisi}) for the sphere. For instance, mimicking the situation
for $\r^{2n},$ one might simply postulate that $\q(f^2) =
\q(f)^2$ for $f \in {\rm su}(2).$ While the squaring rule for angular
momentum is compatible with (\ref{eq:sisi}), one would still ``miss''
various possibilities (corresponding to the freedom in the choice of
parameters $a,\,c$), which do occur in specific
representations.\footnote{\,Because of this,
\cite{KLZ} would refer to (\ref{eq:sisi}) as ``{\em non\/}-Neumann
rules''! }   And in the case of the torus,
\vn\ rules are effectively moot, since the explicit prequantization map
$\q$ itself determines the quantization of every observable. \vn\
rules are also irrelevant in the $\T \r_+$ example, because of the
peculiar structure (\ref{decomp}) of $P(a(1))$. All in all, it appears
as if the \vn\ rules play a secondary role; the basic algebra $\fb$ is
the primary object. It is also more compelling physically and pleasing
{\ae}sthetically to require $\q$ to satisfy an irreducibility
requirement than a \vn\ rule. Still, one can argue that such rules
serve an important purpose \cite{As,Ve}.

There are problems with the polarization approach as well. For one
thing, symplectic manifolds need not be polarizable \cite{go87}. This
rare occurrence not\-withstanding, there are quantizations
which cannot be obtained by polarizing a prequantization: A well-known
example is the extended metaplectic quantization of 
\big(hsp(2$n$,\r),h(2$n$)\big) 
\cite{Bl}. As we shall see presently, the specific predictions of
geometric quantization theory are also off the mark in a number of
instances.

Finally, it should be emphasized that these three approaches to
quantization typically lead to obstructions in one way or another. We
have already seen in \S 5 that \vn\ rules play a crucial role in
deriving the \gr-\vh\ obstructions for
$\r^{2n}$, $S^2$ and $\T\! S^1$. In the context of polarizations, the
only observables which are consistently quantizable {\it ab initio}
are those whose Hamiltonian vector fields preserve a given
polarization
\cite{bl1,Wo}. While this does not preclude the possibility of
quantizing more general observables, attempts to quantize observables
outside this class in specific examples usually result in
inconsistencies. In {\it all\/} instances, the set of {\it a priori\/}
quantizable observables relative to a given polarization forms a
proper Lie subalgebra of the Poisson algebra of the given symplectic
manifold. This observation provides further corroboration that
\gr-\vh\ obstructions to quantization should be the rule rather than
the exception.

\ms

Setting aside the question of the existence of obstructions,
let us now suppose that there is an obstruction to, say, a polynomial
quantization, so that it is impossible to consistently quantize all of
$P(\fb)$. The question is: What are the maximal Lie subalgebras $\oo
\subset P(\fb)$ containing the given basic algebra $\fb$ such that
$(\oo,\fb)$ can be quantized? Modulo technical issues, given a
representation
$\q$ of
$\fb$ on a Hilbert space $\h$, one ought to be able to induce a
representation of its Lie normalizer $\mathfrak n(\fb)$ in $P(\fb)$ on
$\h$. (Indeed, the structure $(\fn(\fb),\fb)$ brings to
mind an infinitesimal version of a Mackey system of imprimitivity
\cite{b-r}.)  Thus it seems reasonable to assert:
\begin{conj} Let $\fb$ be a \fd\ basic algebra. Then every
quantization of
$\fb$ can be extended to a quantization of
$(\fn(\fb),\fb)$.\footnote{\,In \cite{GGT} quantizations which
satisfy this condition are termed ``strong.''} 
\label{conj:obs}
\end{conj}

This is in exact agreement with the examples. In particular, for
$\r^{2n}$ one has $\fn\big({\rm h}(2n)\big) = {\rm hsp}(2n,\r)$, and
for $S^2$ one computes $\fn\big({\rm su}(2)\big) = {\rm u}(2)$.
In both cases, we have shown that these normalizers are in fact the
maximal polynomial subalgebras that can be consistently quantized. 
It is therefore tempting to conjecture that:
\begin{quote}
\emph{No nontrivial quantization
of $(\fn(\fb),\fb)$  can be extended beyond
$\fn(\fb)$.}
\end{quote}

If true,
this would point where to look for a
\gr-\vh\ contradiction, viz. just outside the normalizer. Alas, this
is \emph{false}:  For the cylinder $\fn({\rm e}(2)) = \r \oplus
{\rm e}(2)$. But from \S 5.3, we know that the representation
(\ref{eq:(i)'}) can be extended, in infinitely many ways, to the
quantizations (\ref{eq:qnu}) of $(L^1,P_1)$, where $L^1$
is the Lie subalgebra of observables which are affine in the
(angular) momentum $\ell$. It is not clear how one could
``discover'' this subalgebra given just the basic algebra ${\rm
e}(2)$ (but see below). The situation for $\T \r _+$ is of course
even worse than for $\T \!S^1.$ An outstanding problem is therefore to
determine the maximal Lie subalgebras of quantizable observables.

This is reminiscent of the situation in geometric quantization with
respect to polarizations. Suppose that $\cal A$ is a polarization of
$\ci(M,\!\bc)$. Then one knows that one can consistently quantize
those observables which preserve $\cal A$, i.e., which belong to the
real part of $\fn(\cal A)$ \cite{bl1,Wo}. In this way one obtains a
``lower bound'' on the set of quantizable functions for a given
polarization. If one takes the antiholomorphic polarization on $S^2$,
then it turns out that the set of {\em a priori} quantizable
functions obtained in this manner is precisely the u(2) subalgebra
$\sp\{1,S_1,S_2,S_3\}$. But it may happen that the real part of
$\fn(\cal A)$ is too small, as  for $\r^{2n}$ with the
antiholomorphic polarization. In this case the real part of $\fn(\cal
A)$ is only a proper subalgebra of $P^2$, and in particular is not
maximal. This illustrates the fact, alluded to previously, that the
extended metaplectic representation cannot be derived via geometric
quantization. Furthermore, in the case of the torus, introducing a
polarization will drastically cut down the set of {\em a priori}
quantizable functions, which is at odds with the existence of a full
quantization of this space. So geometric quantization is not a
reliable guide insofar as computing maximally quantizable
Lie subalgebras of observables. On the other hand, the position
subalgebra $S = \{f(q)p + g(q)\}$ (resp. $L^1$) is just the normalizer
of the vertical polarization ${\cal A} = \{h(q)\}$ on $\r^{2}$
(resp. $\{h(\theta)\}$ on $\T \!S^1$), so these subalgebras find
natural interpretations in the context of polarizations.

Clearly, there must be some connection between polarizations and basic
algebras that awaits elucidation. It would be interesting to determine
if there is a way to recast the \gr-\vh\ results in terms of
polarizations. It would also be worthwhile, assuming
that it is somehow possible to predict the maximal set(s) of
quantizable observables {\it a priori},  to see whether one
can use this knowledge to refine geometric quantization theory, or to
develop a new quantization procedure, which is adapted to the \gr-\vh\
obstruction in that it will automatically be able to quantize this
maximal set.

Here we have focused on the quantization of symplectic manifolds. It
is natural to wonder to what extent these results will carry over to
Poisson manifolds, or even to abstract \pa s.

\ms

One of our goals in this paper was to obtain results which are
independent of the particular quantization scheme employed, as long as
it is Hilbert-space based. Therefore it is interesting that some of
the go and no-go results described in this proposal have direct
analogues in deformation quantization theory, since this theory was
developed, at least in part, to avoid the use of Hilbert spaces
altogether
\cite{Bayen e.a.}. So for example,  the no-go result for
$S^2$ is mirrored by the fact that there are no strict SU(2)-invariant
deformation quantizations  of
$\ci (S^2)$
\cite{Ri}, while the go theorem for $T^2$ has as a counterpart the
result that there do exist strict deformation quantizations of the
torus \cite{Ri}. It is generally believed that the existence of
\gr-\vh\ obstructions necessitates a weakening of the Poisson bracket
$\ra$ commutator rule (by insisting that it hold only to order
$\hbar$), but these observations indicate that this may not suffice
to remove the obstructions. There are undoubtedly important
things to be learned by getting to the heart of this analogy. 

\end{section}



\end{document}